\documentclass[referee]{myaa} 
\usepackage{graphicx}
\usepackage{txfonts}

\begin{document}
%
   \title{Abiotic Formation of O$_2$ and O$_3$ in High-CO$_2$ Terrestrial Atmospheres}

   \author{A. Segura \inst{1}\fnmsep\inst{5}
       \thanks{\emph{Present address:} Instituto de Ciencias Nucleares, Universidad Nacional Aut\'onoma
          de M\'exico. Circuito Exterior, C.U. Apartado Postal 70-543, 04510,
          M\'exico, D.F.}
          \and V. S. Meadows \inst{1}\fnmsep\inst{5}
          \and J. F. Kasting \inst{2} \fnmsep\inst{5}
          \and D. Crisp \inst{3}\fnmsep\inst{5}
          \and M. Cohen \inst{4}\fnmsep\inst{5}}
 
  \offprints{A. Segura, \email{antigona@nucleares.unam.mx}}

   \institute{California Institute of Technology, 1200 East California
     Blvd. Mail stop 220-6, Pasadena CA, 91125 USA
     \and Pennsylvania State University, 443 Deike Bldg. State College, PA,
       16802, USA
    \and NASA Jet Propulsion Laboratory, 4800 Oak Grove Dr. Pasadena, CA 91109, USA
     \and University of California, Radio Astronomy Laboratory, 601 Campbell Hall, Berkeley, CA 94720 USA  
     \and Members of the Virtual Planetary Laboratory a project of the NASA Astrobiology Institute}
     
   \date{Received ; Accepted 28 June 2007 }

 
  \abstract
    {Previous research has indicated that high amounts of ozone (O$_3$) and oxygen (O$_2$) may be
    produced abiotically in atmospheres with high concentrations of CO$_2$.  The abiotic production
    of these two gases, which are also characteristic of photosynthetic life processes, could pose
    a potential ``false-positive'' for remote-sensing detection of life on planets around other
    stars. We show here that such false positives are unlikely on any planet 
    that possesses abundant liquid water, as rainout of oxidized species onto a reduced planetary
    surface should ensure that atmospheric H$_2$ concentrations remain relatively high, and that
    O$_2$ and O$_3$ remain low.}
   {To determine the amount of O$_3$ and O$_2$ formed in a high CO$_2$
    atmosphere for a habitable planet without life.}
   {We use a photochemical model that considers hydrogen (H$_2$) escape and a detailed
   hydrogen balance to calculate the O$_2$ and O$_3$ formed on planets with 
   0.2 of CO$_2$ around the Sun, and 0.02, 0.2 and 2 bars of
   CO$_2$ around a young Sun-like star with higher UV radiation. The
   concentrations obtained by the photochemical model were used as input in a 
   radiative transfer model that calculated the spectra of the modeled planets.}   
  {The O$_3$ and O$_2$ concentrations in the simulated planets are extremely small, and unlikely
  to produce a detectable signature in the spectra of those planets.}
   {With a balanced hydrogen budget, and for planets with an
  active hydrological cycle, abiotic formation of
  O$_2$ and O$_3$ is unlikely to create a possible false positive for life detection in either the 
  visible/near-infrared or mid-infrared wavelength regimes.}


\titlerunning{Abiotic O$_2$ and O$_3$ in high CO$_2$ atmospheres}
\authorrunning{Segura et al.}

   \maketitle

\section{Introduction}

Most of the O$_2$ in Earth's present atmosphere is thought to have been produced by oxygenic photosynthesis,
followed by burial of organic carbon in marine sediments (Cloud \cite{cloud}; Walker \cite{walker}; Holland 
\cite{holland78}, \cite{holland84}, \cite{holland}). 
Prior to the origin of life, and of O$_2$-producing life in particular, atmospheric O$_2$ mixing ratios
are thought to have been very low, $10^{-13}$ by volume, or $\sim 10^{-12}$ PAL at the surface (Walker \cite{walker}; 
Kasting et al. \cite{kastingetal79}; Kasting \cite{kasting93}; Kasting \& Catling \cite{kascat}) Here, `PAL' 
means `times the Present Atmospheric Level', which is
21 percent by volume, or 0.21 bars. O$_2$ concentrations in the upper atmosphere of the early Earth could 
have been much higher, up to $\sim 10^{-3}$ by volume (Kasting and Catling \cite{kascat}), 
as a consequence of photolysis of CO$_2$,
followed by recombination of O atoms to make O$_2$. A small amount of ozone, O$_3$, could have been formed
from this O$_2$, but not enough to provide an effective shield against solar UV radiation (Levine et al. 
\cite{levineetal};
Kasting and Donahue \cite{kasting80}; Kasting \cite{kasting93}, and refs. therein; Cockell and Horneck,
\cite{CockellHorneck}). Early life therefore 
probably evolved in an anoxic, high-UV environment.

While the question of abiotic O$_2$ and O$_3$ levels has historically been of interest to geologists 
and biologists, it has recently become an important issue for astronomers as well. Within the 
next 15-20 years, space-based telescopes, such as the two Terrestrial Planet Finder (TPF) 
missions planned by NASA (http://planetquest.jpl.nasa.gov/TPF/tpf\_c\_team.cfm) and ESA's 
Darwin mission (http://ast.star.rl.ac.uk/darwin/), will hopefully search for Earth-sized planets around 
other stars and attempt to obtain spectra of their atmospheres and surfaces. TPF-C (a visible/near-IR coronagraph) 
will be sensitive to the 0.76-$\mu$m band of O$_2$, while TPF-I and Darwin (both thermal-IR interferometers) 
will be sensitive to the 9.6-$\mu$m band of O$_3$. Both of these gases are considered to be possible 
biomarker compounds (Owen \cite{Owen}; Angel \cite{angeletal}; Leger et al. \cite{legeretal}). The question 
as to whether they can be produced abiotically is therefore of great potential 
relevance to the interpretation of data from these future missions.

Situations in which either O$_2$ or O$_3$, or both, might accumulate abiotically have been identified by a number of
different authors. Some of these situations are legitimate ``false positives'' for life, while others might be 
misleading. Two situations, in particular, appear capable of producing ``false positive'' signals. 
The first is a runaway greenhouse planet, like early Venus, on which large amounts of hydrogen 
escape from a hot, moist atmosphere (Kasting \cite{kasting97}; Schindler \& Kasting \cite{SchindlerKasting2000}). 
Because the hydrogen originates from H$_2$O, oxygen is left behind. The escape of a
terrestrial ocean equivalent of hydrogen, unaccompanied by oxygen sinks, could leave an atmosphere containing $\sim 240$  
bars of O$_2$ (Kasting \cite{kasting97}). A second conceivable ``false positive'' is a frozen, 
Mars-like planet that is large enough to retain heavy gases but too small to maintain volcanic outgassing 
(Kasting \cite{kasting97}; Schindler \& Kasting \cite{SchindlerKasting2000}). The frozen surface
would inhibit the loss of oxygen by reaction with reduced minerals, whereas the lack of outgassing would eliminate reaction 
with reduced volcanic gases (primarily H$_2$) as an oxygen sink. The martian atmosphere contains $0.1 \%$
O$_2$ and would likely have even more if the planet were slightly larger so that it did not lose oxygen to space by nonthermal
loss mechanisms (McElroy and Donahue \cite{mcelroy}).

Both of the ``false positives'' mentioned above apply to planets that lie outside of the liquid water habitable
zone around their parent star.  The boundaries of this zone can be estimated to first order from climate models (Kasting et al.
\cite{kastingetal93}), and it should be possible to determine observationally whether the planet's distance from its parent
star falls outside these limits. Such a determination would not be definitive, as the theoretical limits of the
habitable zone are uncertain, mostly because of the difficulty in simulating clouds. (The Kasting et al. 
\cite{kastingetal93} model effectively
puts the cloud layer at the planet's surface, thereby ignoring all cloud
feedbacks.)
Planets outside
the habitable zone may also be distinguishable by the absence, or near
absence, of gaseous H$_2$O in their spectra, although water-rich planets
near the inner edge might not obey this rule, and one should remain
suspicious of O$_2$ signals on such bodies.
Still, the most important false positive issue is whether or not planets
within the habitable zone could build up O$_2$ or O$_3$
abiotically.

Calculations predicting high, or relatively high, abiotic O$_2$ and O$_3$ concentrations 
have appeared sporadically in the literature over the past 40 years. All authors have realized that photolysis of 
H$_2$O, followed by escape of hydrogen to space, is a net source of oxygen. Berkner and 
Marshall (\cite{berkner64}, \cite{berkner65}, \cite{berkner66}, \cite{berkner67}) estimated abiotic O$_2$
concentrations of 10$^{-4}$ - 10$^{-3}$ PAL based on how much O$_2$ buildup was 
needed to block out the UV radiation that dissociates H$_2$O. This is probably 
too little O$_2$ to be detectable spectroscopically by a telescope
like TPF-C; however, the associated ozone layer could conceivably be detected by a telescope like TPF-I or
Darwin (Segura et al. \cite{seguraetal03}).
In a study a few years later, Brinkman (\cite{brinkman}) found abiotic O$_2$ concentrations
of up to 0.27 PAL. If correct, such an atmosphere would produce an absorption feature at 0.76 $\mu$m about
half as strong as that of Earth (Des Marais, et al. \cite{desmaraisetal}). Brinkmann obtained this high value 
because he assumed that precisely one-tenth of the H atoms 
produced by H$_2$O photolysis escaped -- a fraction that we now know is much too high.

All of these
early studies were performed before the factors controlling hydrogen escape from Earth's atmosphere were well
understood. Hunten (\cite{hunten}) showed that the H escape rate from Earth (and from Saturn's moon, Titan) is 
limited by the rate at which hydrogen can diffuse upwards through the homopause. (The homopause, 
near 90 km in Earth's atmosphere, is the altitude above which light gases begin to separate 
out from heavy gases. Equivalently, it marks the transition from turbulent vertical transport 
to molecular diffusion.) The resulting escape rate is termed the ``diffusion-limited
flux''. Hydrogen can escape more slowly than the diffusion limit because of energy limitations 
higher up (see, e.g., Tian et al. \cite{tianetal}), but it cannot escape more rapidly. Walker (\cite{walker}) applied this concept to 
the early Earth and showed that it implies extremely
low ground-level prebiotic O$_2$ concentrations, of the order of 10$^{-13}$ PAL. This same reasoning applies to
vertically resolved atmospheric photochemical models (Kasting et al. \cite{kastingetal79}; Kasting \cite{kasting93}; 
Kasting and Catling \cite{kascat}), although the
O$_2$ concentration in the stratosphere can be much higher ($10^{-3}$), as
mentioned earlier. 
High stratospheric O$_2$
concentrations are favored by high CO$_2$ levels and high solar UV fluxes (Canuto et al. \cite{canutoetal82},
 \cite{canutoetal83}). It should be pointed out that the extremly high UV fluxes studied by Canuto et al. are
unlikely to apply to the early Earth or to any analogous planet, as the T-Tauri phase of stellar evolution was probably
over long before the Earth was fully formed.

Although most recent models predict low abiotic O$_2$ concentrations for planets within the habitable
zone -- contradictory results also appear. This happens,   
for example, in some calculations performed by Selsis et al. (\cite{selsisetal02}) 
for a variety of Earth-like and Mars-like planets. The Earth-like planets all had significant
volcanic sinks for oxygen, and so none of these cases produced high O$_2$ or O$_3$. However, 
their ``early Mars-type'' planet (Case B2), which assumed zero volcanic outgassing, exhibited 
0.1 PAL of O$_2$ and a ``super'' ozone layer with a column
depth of $\sim 0.7$ atm cm -- roughly twice that of modern Earth. 
This model simulated a warm, humid, 1-bar CO$_2$ atmosphere underlain by an ocean. 
If this calculation were valid, then such a planet would represent
another possible ``false positive'' for life. Selsis et al. argued that the O$_3$ 9.6-$\mu$m 
band would be obscured by the neighboring 9.4- and 10.4-$\mu$m hot bands of CO$_2$, 
and thus would not constitute a false positive for Darwin
(or TPF-I). However, the 0.76-$\mu$m band of O$_2$ would be quite prominent in this atmosphere, 
and so it would remain an issue for TPF-C.

We argue here that the Selsis et al. calculation is incorrect, or at least over-simplified, and that this
case is not a likely ``false positive''. More precisely, we suggest that abiotic O$_2$ buildup is only likely 
to occur on dry or frozen planets, and such planets could be differentiated from warmer Earth-like planets
by examining their spectra. The reason that O$_2$ reached such high levels in the Selsis et al. calculation
is that the authors failed to
consider the effect of rainout of oxidized (or reduced) species on the atmospheric hydrogen budget.
Photochemical production of oxidized species such as hydrogen peroxide, H$_2$O$_2$, followed by their reaction with
reduced species in the crust or in seawater, should generate a net source of H$_2$ (Kasting et al. 
\cite{kastingetal84}; Kasting and Brown \cite{kasbrown98}; Kasting and Catling \cite{kascat}). 
This H$_2$ source (and O$_2$ sink) was left out of the Selsis et al. model, allowing for a much higher
abiotic O$_2$ concentration. The next section describes
the atmospheric hydrogen budget, along with the rest of 
our photochemical model, in more detail. We then use our model to simulate
various Earth-like planets, with and without volcanic outgassing, and under high 
and low stellar UV flux, 
and we calculate abiotic O$_2$ and O$_3$ 
concentrations and accompanying visible and thermal-IR spectra. Our goal is to determine whether 
any additional ``false positives'' for O$_2$ producing life could exist in high-CO$_2$ atmospheres.

\section{Model description}

\subsection{The photochemical model}
The 1-D photochemical model used in our study was developed for high-CO$_2$/high-CH$_4$ 
terrestrial atmospheres by 
Pavlov et al. (\cite{pavlov}) and was subsequently modified by Kharecha et al. (\cite{kharecha}). The model 
simulates an anoxic atmosphere composed of 0.8 bar of N$_2$ and variable amounts of CO$_2$. We did not simulate
a pure-CO$_2$, ``Mars-like'' atmosphere explicitly; however, such an atmosphere is similar in principle
to our high-CO$_2$ cases.
Our photochemical model contained 73 chemical species involved in 359 reactions and spanned the region from the
planetary surface up to 64 km in 1-km steps. The solar zenith angle was fixed
at 50\degr , and a two-stream approach was used for the radiative transfer. The continuity equation was solved at each height for each of 
the long-lived species, including transport by eddy and molecular diffusion. The combined equations
were cast in centered finite difference form. Boundary conditions for each species were applied at the top
and botton of the model atmosphere, and the resulting set of coupled differential equations was 
integrated to steady state using the reverse Euler method. Two different stellar UV fluxes were used, as described below. The eddy 
diffusion profile for most cases was assumed to be the same as the one measured on present Earth (Massie \& Hunten 
\cite{MasHun}), with modifications for atmospheres in which the total surface pressure exceeded one bar. All of the simulated planets were assumed to be devoid of life; hence, none of the compounds in the atmosphere 
was considered to have a biological source.

\begin{figure}
   \resizebox{\hsize}{!}{\includegraphics{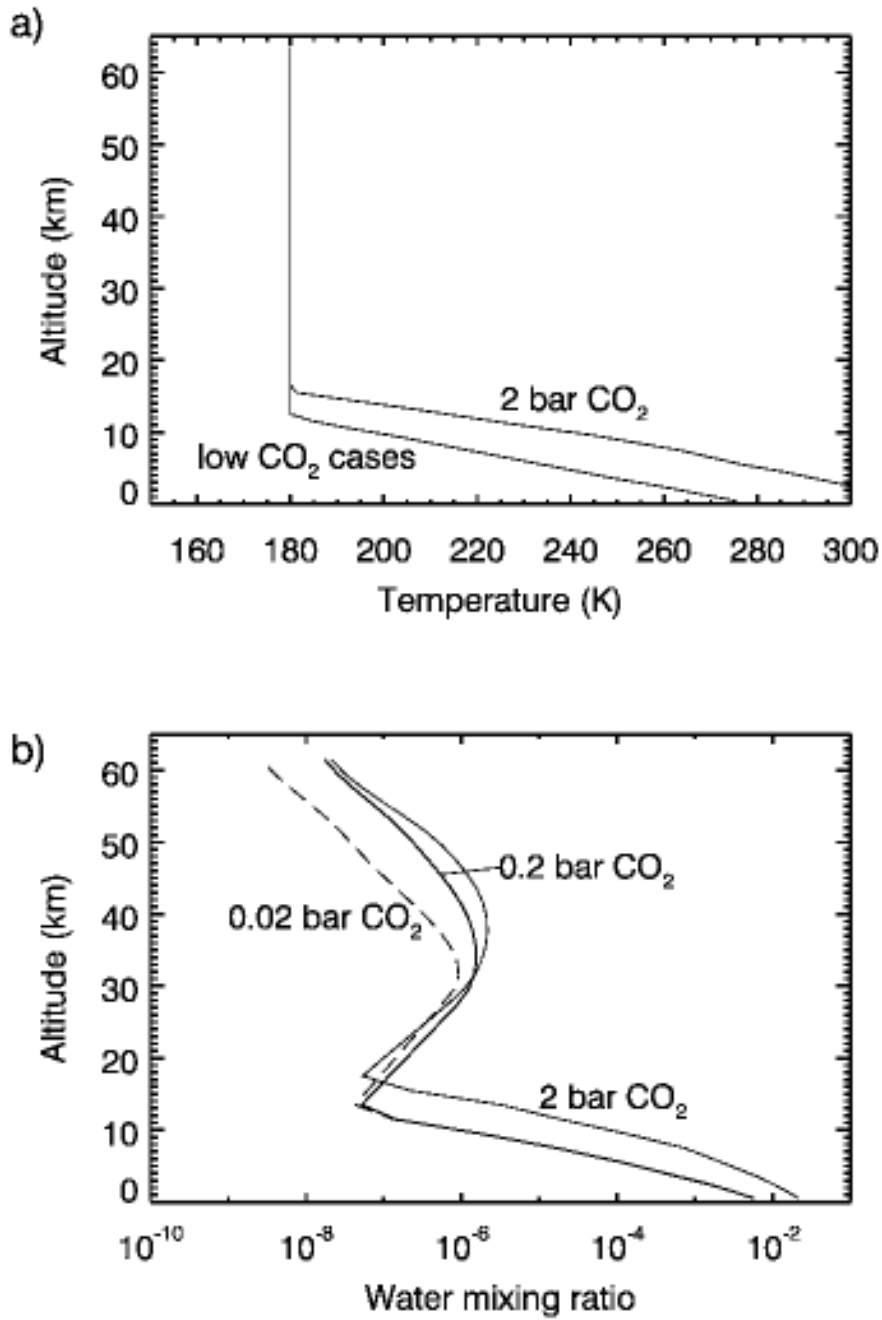}}
      \caption{a) Temperature profiles for the low-CO$_2$ atmospheres (0.2 and 0.02 bars of CO$_2$)
         and the high CO$_2$ planet (2 bars of CO$_2$.
         b) Water profiles for the simulated planetary atmospheres around EK Dra. }
         \label{profiles}
   \end{figure}

As a reference, or ``standard'', case we assumed a 1-bar, cloudless, ``early-Earth'' type atmosphere 
with 0.2 bar of CO$_2$. 
 This is roughly the amount of CO$_2$ needed to compensate for a 30\% reduction in solar luminosity, appropriate for a 
time near 4.5 Gyr ago.
(Kasting \cite{kasting93}). We did not perform coupled climate model calculations for this case; rather, we simply
assumed that the temperature decreased with height from 278 K at the surface to 180 K at 12.5 km, following a
moist adiabat. Above that height, the atmosphere was assumed to be isothermal (Fig. \ref{profiles}). 
This is roughly consistent with the predictions of climate models (Kasting \cite{kasting90}). It is difficult for us
to improve significantly on this assumption, as our own climate model (Pavlov
et al. \cite{pavlov2000}) is not particularly
accurate up in the Doppler-broadened stratosphere and mesosphere. The assumed
solar UV flux in this model was the same as for Earth today. We also performed simulations for higher UV fluxes.
As discussed further below, stellar UV fluxes
at wavelengths less than $\sim$200 nm originate from a star's chromosphere and are, if anything, anti-correlated
with the flux at longer wavelengths. Stars tend to brighten in the visible as they age; however, their short-
wavelength UV emissions decrease with age in most stars because their spin rate slows and they exhibit less 
chromospheric activity. Longer wavelengths are included in the model, but they mostly affect the temperature 
profile, which has been fixed for the planets simulated here. 

\begin{figure}
   \resizebox{\hsize}{!}{\includegraphics{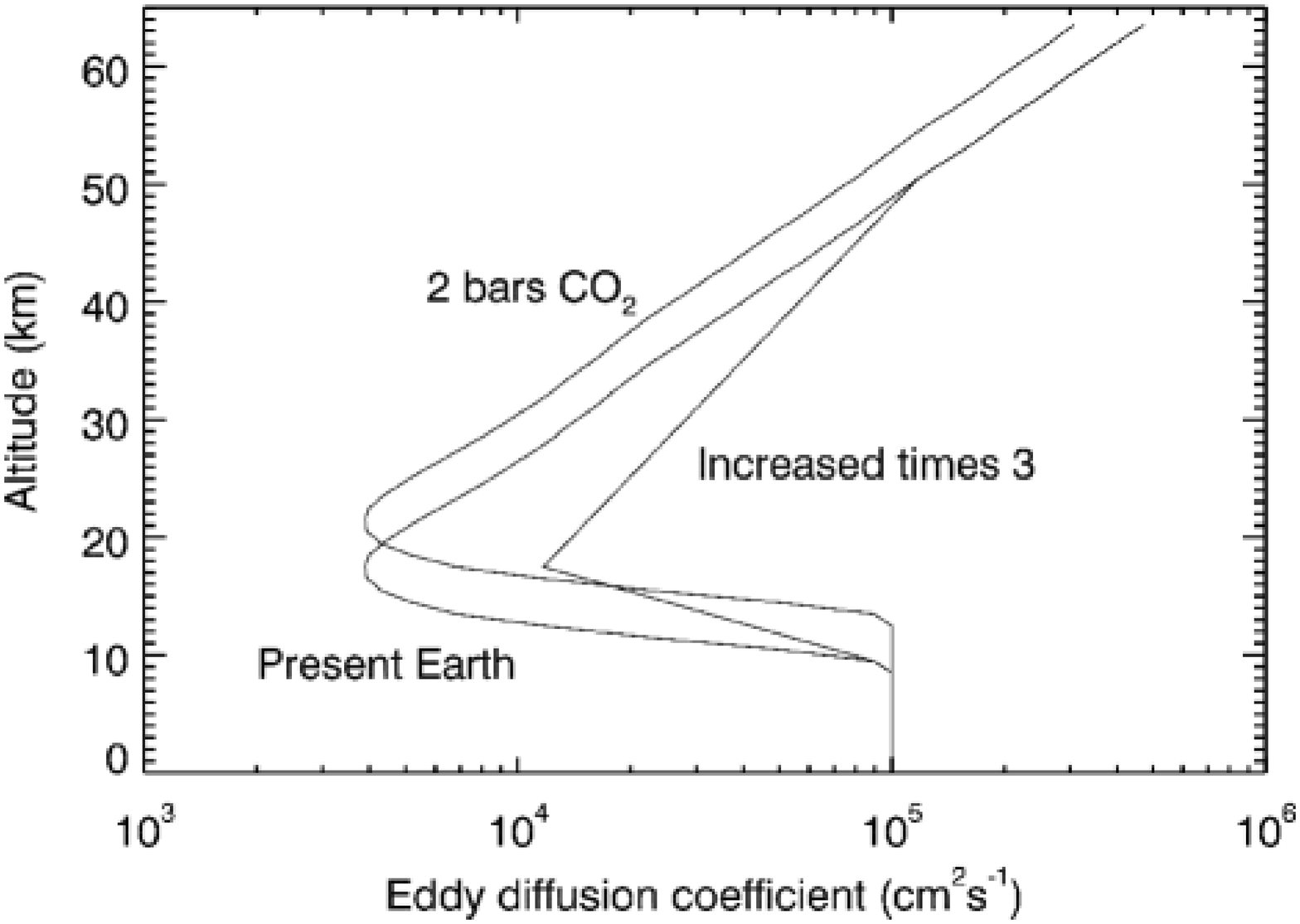}}
      \caption{Eddy diffusion coefficient ($k_{\mathrm{edd}}$) profiles used in our
        simulations. Present Earth $k_{\mathrm{edd}}$ is the one measured by
        Massie \& Hunten (\cite{MasHun}), the other two profiles are modified versions of this
        coefficient for the 2 bar of CO$_2$ atmosphere and more pausible $k_{\mathrm{edd}}$ for early Earth (see Results).}
         \label{eddy}
   \end{figure}

We also did calculations for early Earth-type planets with CO$_2$ partial pressures of 0.02, 0.2, and 
2 bars and higher UV radiation than the present solar flux. For the 
0.02-bar case, we kept the
surface temperature and pressure the same as for the standard model. (This implies slightly more N$_2$ and
either a higher incident solar flux or additional unspecified greenhouse warming.) For the 2-bar CO$_2$ case, the
total surface pressure was 2.9 bars (= pCO$_2$ + pN$_2$ + pH$_2$O). The surface temperature, 317 K, and the vertical
temperature profile for this atmosphere were calculated by a climate model (Kasting \cite{kasting90}). In this latter
case, the eddy diffusion profile was readjusted to account for the increased troposphere thickness. Following
Kasting (\cite{kasting90}) we shifted the eddy diffusion profile upward by 4 km for the 2 bar case, and assigned
an eddy diffusion coefficient of $10^5$ cm$^2$ s$^{-1}$ to the 4 lowermost
grid points (Fig \ref{eddy}). Although the cases that we
studied are not entirely self-consistent, we chose them because they span a plausible range of atmospheric 
composition for abiotic, early Earth-type planets (Kasting \cite{kasting90}). 

Volcanic outgassing of H$_2$ and other reduced species was included in most, but not all, of the model
calculations. For H$_2$, we used the present volcanic outgassing rate $\sim 5\times 10^{12}$ mol yr$^{-1}$, or 
$\sim 2\times 10^{10}$ cm$^{-2}$ s$^{-1}$ (Holland \cite{holland}). Sleep and Bird (\cite{SleepBird})
argue for an H$_2$ outgassing rate that is lower by a factor of 10. Our calculations span this range because
we consider zero outgassing as well.
The methane flux for the standard case was set to the estimated modern non-biogenic flux on hydrothermal vents.
Formerly, the biotic-to-abiotic CH$_4$ ratio was estimated to be $\sim 300:1$ (Kasting \& Catling \cite{kascat}).
New measurements of methane dissolved in the fluids of the Lost City vent field (Kelley \cite{kelley}) indicate 
that the abiotic methane flux is higher than previously thought by a factor of 10; therefore, the biotic-to-abiotic
ratio may be $\sim 30:1$. The present biological methane flux is 535 Tg CH$_4$/year or $3.3 \times 10^{13}$ mol/yr 
(Houghton et al. \cite{houghton}), so the assumed abiotic methane flux was $1 \times 10^{12}$ mol/yr, or
$3.7 \times 10^9$ cm$^{-2}$s$^{-1}$. The actual abiotic methane flux may be lower than this value 
if the CH$_4$ in
the vent fluids is produced by methanogens living deep within the vent systems. An (implicit)
NH$_3$ flux of $3.87\times 10^{9}$ cm$^{-2}$ s$^{-1}$ was included because the standard model contained 10 ppbv
of NH$_3$. This was simply for convenience, because NH$_3$ was included in the species list and, hence, its 
concentration must be non-zero.
For the ``no-outgassing'' case, the methane and H$_2$ fluxes were set equal to zero, 
but volcanic outgassing of SO$_2$ 
and a tiny amount of NH$_3$ was retained in order to avoid numerical problems associated with 
disappearing species within the code. The implications of this assumption are discussed below.

Our model also includes the volcanic outgassing of SO$_2$ at its present estimated release rate 
$3.5$~$\times$~$10^9$~cm$^{-1}$~s$^{-1}$
(Kasting \cite{kasting90}) along with formation of NO, O$_2$ and CO by lightning in the troposphere, as described by 
Kasting (\cite{kasting90}).Rainout rates for soluble species are calculated using the method of Giorgi \& Chameides
(\cite{giorcham}). These rainout terms are added to the normal
chemical loss rate for each species.  Sulfur gases are removed from the model atmosphere by rainout
  and surface deposition and by conversion into particulate sulfur and
  elemental sulfur, followed by rainout and surface deposition of the
  particles. Rainout lifetimes of high soluble gases ({\it e.g.} H$_2$SO$_4$)
  and of particles was approximately five days. Less soluble gases (SO$_2$,
  H$_2$S) have longer lifetimes against rainout, according to this model. The
  lifetimes assumed here are appropiate for the modern atmosphere. we assume
  that the hydrological cycle and, hence, rainout lifetime was similar on
  early Earth. This assumption seems reasonable, since evaporation (and, thus,
precipitation) rates are ultimately controlled by solar heating rates, not by
temperature (Holland, \cite{holland78}). Furthermore, the rainout rate of highly soluble
gases is determinated by how often it rains, no how much it rains.
Soluble species are also removed by direct deposition at the lower boundary,
simulating uptake by the ocean. The downward flux of a chemical species $i$ is equal to its number density, $n_i$,
multiplied by a deposition velocity, $v_\mathrm{dep}$. Deposition velocities for different gases are taken 
from Slinn et al. 
(\cite{slinnetal}) and Lee \& Schwartz (\cite{lee}). Values range from $\sim 0.02$ cm/s for less reactive species 
(e.g., H$_2$O$_2$) to 1 cm/s for the most reactive species (e.g., OH). The upper limit on $v_\mathrm{dep}$ is set
by diffusion through the turbulent atmospheric boundary layer, assuming that molecules are absorbed by the surface
each time they collide with it.

One may ask how our photochemical model achieves steady state when species are continually injected into the
atmosphere from volcanoes. For hydrogen and sulfur, the answer is clear: they are removed either by escape to
space (for H) or by rainout from the atmosphere. For CH$_4$ and NH$_3$, the answer is less obvious. These species are
oxidized photochemically to CO$_2$ and N$_2$, respectively. As the CO$_2$ and N$_2$ concentrations are held constant
in the model atmosphere, these species are implicitly lost by reactions of CO$_2$ with surface minerals (to form
carbonates) and by unspecified mechanisms of N$_2$ loss (most plausibly biological nitrogen fixation).

\subsection{The atmospheric hydrogen budget}
A key feature of our model that makes it useful for this study is its ability to keep track of the atmospheric 
hydrogen budget, or redox budget. The basic principle behind this budgeting scheme is simple: when one species
is oxidized, another species must be reduced, and vice versa. Alternatively,
one could rephrase this principle as requiring that the electron budget of the model atmosphere be balanced.

In practice, it is easiest to keep track of redox balance in terms of the abundance of molecular hydrogen, H$_2$.
Following Kasting and Brown (\cite{kasbrown98}) we define ``redox-neutral'' species: 
H$_2$O (for H), N$_2$ (for N), CO$_2$ (for C), and SO$_2$ (for S). 
All other species are assigned redox coefficients relative
to these gases by determining how much H$_2$ is produced or consumed during their formation from redox neutral
species. For example, formation of hydrogen peroxide can be expressed as:

\noindent $2 \mathrm{H}_2\mathrm{O} \longrightarrow \mathrm{ H}_2\mathrm{O}_2 + \mathrm{H}_2$
       
The coefficient of H$_2$O$_2$ in the hydrogen budget is therefore $-1$, meaning that when one H$_2$O$_2$ molecule
is rained out, one H$_2$ molecule is produced. Similarly, formation of elemental sulfur, S$_8$, can be written
as:

\noindent $8 \mathrm{SO}_2 + 16 \mathrm{H}_2 \longrightarrow \mathrm{ S}_8 + 16 \mathrm{H}_2\mathrm{O}$

Hence, rainout of S$_8$ particles consumes 16 H$_2$ molecules (per S$_8$ molecule), so the redox coefficient
of S$_8$ is $+16$. Volcanic outgassing of species other than H$_2$ itself is treated in the same manner, {\it i.e.},
it adds to the total hydrogen outgassing rate according to the appropriate stoichiometry. As mentioned earlier,
the standard model includes modest outgassing of CH$_4$ and NH$_3$, in addition to H$_2$. 
Thus, the total hydrogen outgassing rate can be written as:

\noindent $\Phi_{\mathrm{volc}}(\mathrm{H}_2) = \Phi(\mathrm{H}_2) + 1.5\Phi(\mathrm{NH}_3) + 4\Phi(\mathrm{CH}_4)$

\noindent where the terms on the right represent the fluxes of the individual reduced species multiplied by 
their stoichiometric coefficients in the redox budget

With these definitions in place, the hydrogen budget can be written as:

   \begin{equation} \label{hbal}
      \Phi_{\mathrm{volc}}(\mathrm{H}_2) + \Phi_{\mathrm{rain}}(\mathrm{Ox}) = 
        \Phi_{\mathrm{esc}}(\mathrm{H}_2)  +
        \Phi_{\mathrm{rain}}(\mathrm{Red}) 
   \end{equation}

Here, the terms $\Phi_\mathrm{rain}$(Ox) and $\Phi_\mathrm{rain}$(Red) represent the net flux of photochemically produced
oxidants and reductants, respectively, from the atmosphere to the ocean. This flux includes both rainout
and surface deposition. Eq.~(\ref{hbal}) is diagnostic, not prognostic; hence, it provides a good check both on
redox balance and on the photochemical scheme. (This equation will not balance if any of the chemical
reactions are not balanced.) In our model, Eq.~(\ref{hbal}) balances to within one part in $10^3$ for the standard case and 
one part in $10^5$ for the other simulated planets (Table \ref{H2budget}).

As discussed earlier, we assume that the hydrogen escape is limited by the H$_2$ diffusion rate through the
homopause. The diffusion-limited escape flux is given by Hunten (\cite{hunten}) and Walker (\cite{walker}):

  \begin{equation} \label{phiesc}
    \begin{array}{ll}
      \Phi_{\mathrm{esc}}(\mathrm{H}_2) & =  \left(\frac{b}{H} \right) f_{\mathrm{tot}} \\
                                        & \cong 2.5 \times 10^{13} f_{\mathrm{tot}}
                                          \mbox{ (molecules cm$^{-2}$ s$^{-1}$) } 
    \end{array}
  \end{equation}
where $b$ is an average binary diffusion coefficient for the diffusion of H
and H$_2$ in nitrogen, $H$ is the atmospheric scale height $(RT/g)$, and $f_{\mathrm{tot}}$ is the sum of the mixing ratios 
of all the hydrogen-containing species, weighted by the number of hydrogen atoms they contain.
As we have expressed the hydrogen budget in terms of H$_2$ molecules, we do the same for $f_{\mathrm{tot}}$:
   \begin{equation} \label{ftot} 
      f_{\mathrm{tot}} = f(\mathrm{H}_2) + 0.5f(\mathrm{H}) + f(\mathrm{H}_2\mathrm{O}) 
                + \ldots     
   \end{equation}

Because the stratosphere is cold and dry in all of our simulations, the contribution of H$_2$O to 
f$_{\mathrm{tot}}$(H$_2$)
is negligible. When this is not the case (as in the runaway greenhouse example mentioned earlier), high abiotic
O$_2$ concentrations are possible. The actual hydrogen escape rate may be lower than predicted by Eq.~(\ref{phiesc}) 
However, that is not a concern for this paper, as slower H escape will lead to higher atmospheric H$_2$ 
concentrations, and hence lower abiotic O$_2$ and O$_3$. When trying to identify ``false positives'' for life,
the conservative approach is to assume the maximum, diffusion-limited, hydrogen escape rate.

\subsection{The input stellar spectra}
Some of the stars observed by the TPF missions and by Darwin may be young stars with UV fluxes higher than that
of the Sun. High stellar UV should increase the rate of CO$_2$ photolysis, and thus enhance high-altitude 
concentrations of O$_2$, and possibly O$_3$, on planets
orbiting such stars. High atmospheric CO$_2$ concentrations on young,
early Earth-type planets would presumably enhance production of O$_2$ and O$_3$. To determine whether this process 
might produce enough abiotic O$_2$ and O$_3$ to be detected remotely, we simulated planets 
with high-CO$_2$ atmospheres around a star with a much higher UV
flux than our Sun. EK Draconis was chosen from a suite of solar-type
stars studied by Ribas et al. (\cite{ribas}) because it has the highest UV flux of
the six stars studied.  EK Dra itself may be an unlikely candidate for study by TPF or Darwin 
because of its extremely young age (see Discussion). However, our goal here was to study a planet exposed to
a maximal UV radiation flux to determine if this environment favors abiotic formation of O$_2$ and O$_3$.
EK Dra (HD 129\,333) is a G1.5 V spectral type 
member of the Pleiades moving group. It has an estimated age of $\sim$ 0.1 Gyr, making it a 
good proxy for the young Sun (Dorren \& Guinan \cite{dorren}; Strassmeier \&
Rice \cite{strassmeier};
 Montes et al. \cite{montes}; Ribas et al. \cite{ribas}).
Dorren \& Guinan (\cite{dorren}) state that it is the ``only known star 
closely matching the zero age main sequence Sun observable with IUE'' (the International 
Ultraviolet Explorer.)

An absolute stellar spectral energy distribution was assembled from a
 Kurucz (\cite{kurucz}) photospheric spectrum with $T_{\mathrm{eff}}=5765$~K, 
surface gravity $\log g=4.61$~g~cm$^{-2}$, and solar
metallicity. (Gray et al. (\cite{gray}) derived [M/H]
$-0.04 \pm 0.09$ dex).  The spectrum was absolutely normalized at 21 observed
photometric points between 0.38 and 12 $\mu$m, as described by Cohen et al. (\cite{cohen}).  
 Nineteen  empirical UV spectra (17 SW, 2 LW) were recalibrated
(Massa \& Fitzpatrick \cite{massa}) and combined using the overlap.  From 0.335
to 160 $\mu$m the spectrum is the normalized photosphere.  Below 0.335 $\mu$m
it is an 11-point boxcar-smoothed version of the empirical UV data.

\begin{figure}
   \resizebox{\hsize}{!}{\includegraphics{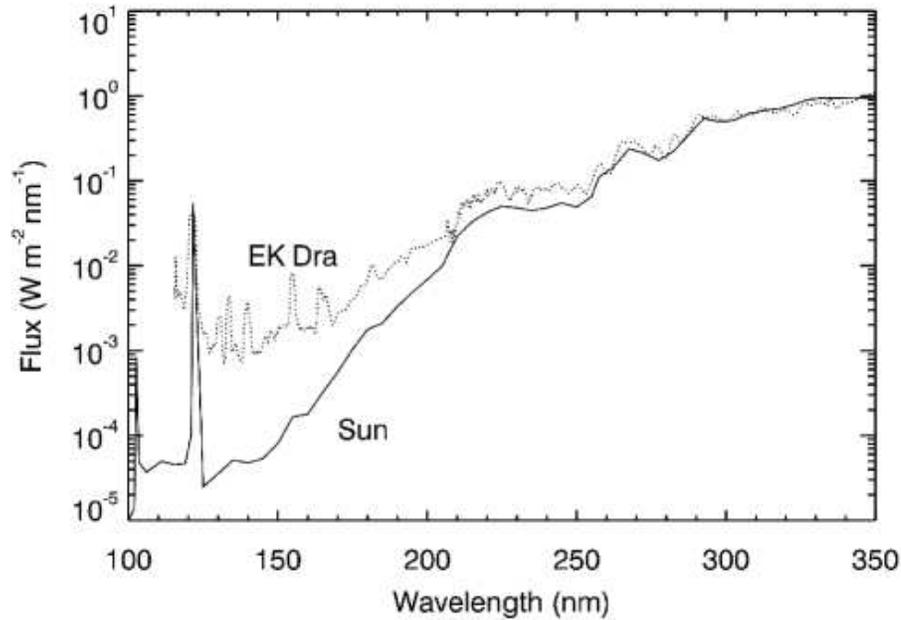}}
      \caption{UV fluxes for the Sun and EK Dra at the top of the atmosphere of a
               habitable planet at 1 AU equivalent distance.}
         \label{uvflux}
   \end{figure}

The Ly$\alpha$ flux has not been measured for EK Dra. Therefore, we estimated its flux
using the empirical expression derived by Ribas et al. (\cite{ribas}, Eq. (2)). At the top of the planet's
atmosphere, the Ly$\alpha$ flux was estimated to be 105.8 ergs cm$^{-2}$ s$^{-1}$. By comparison, the Ly$\alpha$ 
flux from our Sun varies from $\sim$ 6 to 10 ergs cm$^{-2}$ s$^{-1}$ during one 
activity cycle at 1 AU (Emerich et al. \cite{Emerichetal2005}). The simulated planets around 
EK Dra were positioned so as to receive the same integrated solar flux as present Earth (1375 W m$^{-2}$).
This condition is satisfied if the planet's orbital radius is 0.97 AU. (EK Dra is slightly less luminous
than the present Sun.) The UV flux at the top 
of the simulated planet's atmosphere is shown in Fig.~\ref{uvflux}. Earth's UV flux is also shown for comparison.
The solar flux used in the code comes from the World Meteorological Organization (\cite{WMO})

\subsection{Radiative transfer model}

To generate the spectra shown in this paper, the pressure and
temperature profiles and calculated mixing ratios from the coupled
climate-chemical model are used as input to a line-by-line radiative
transfer model, which generates angle-dependent synthetic radiance
spectra.   As the first step in this process the temperature and
pressure profiles, and the calculated mixing ratios for H$_2$O,
O$_2$, O$_3$, CO$_2$, and CH$_4$ are input to a model (LBLABC:
Meadows and Crisp \cite{MeadowsCrisp96}) which generates 
line-by-line monochromatic gas absorption coefficients. LBLABC employs several nested
spectral grids that completely resolve the narrow cores of individual
gas absorption lines at all atmospheric levels, and include their
contributions at very large distances from the line center (e.g. 500
to 1000 cm$^{-1}$ from the line centers.) 

This model was originally developed for simulating radiative processes in the deep atmosphere of Venus 
(Meadows and Crisp \cite{MeadowsCrisp96}) and for highly precise simulation of 
radiances and heating rates in the Earth's atmosphere (Crisp \cite{crisp97}).
In addition, since we are
working with relatively massive CO$_2$ atmospheres, we explicitly
included the effects of collisionally-induced absorption for CO$_2$.
Our current models include CO$_2$ continuum absorption at all 
wavelengths where it has been measured or predicted.  At wavelengths between 
33 and 250 $\mu$m, we use the pressure induced rotational absorption (PIA)
formulation described by Gruszka and Borysow (\cite{GruzkaBorysow96},
\cite{GruzkaBorysow97}, \cite{GruzkaBorysow98}).  At wavelengths between 4 and
10 $\mu$m, we simulate the CO$_2$ PIA using the formulation described by
Kasting et al. (\cite{kastingetal84b}).  At near-IR wavelengths, we used the empirical pressure-induced absorption coefficients from Moore (\cite{moore}).  

LBLABC simulates the line shape function differently in the
 line-center and far-wing regions. For line-center distances less than 40
 Doppler halfwidths, a Rautian line shape is used.  This line shape
 incorporates Doppler broadening, collisional (Lorentzian) broadening, 
and collisional (Dicke) narrowing (c.f. Goody and Yung \cite{GoodyYung89}).  
At greater distances, a Van Vleck-Weisskopf profile is used for 
all gases except for H$_2$O, CO$_2$, and CO.  The super-Lorentzian 
behavior of the far wings of H$_2$O lines, which has been attributed to 
the finite duration of collisions, is parameterized by multiplying the
 Van Vleck-Weisskopf profile by a wavenumber-dependent X factor 
(Clough et al. \cite{cloughetal}).  This profile is adequate to account for the 
water vapor continuum absorption seen at thermal infrared wavelengths, 
as well as the weak continuum absorption seen throughout much of the 
visible and near-infrared spectrum (Crisp \cite{crisp97}).  The sub-Lorentzian 
behavior of the CO$_2$ line profile is primarily a consequence of a 
vibration-rotation energy redistribution process called collisional 
line mixing.  Direct ab initio methods now exist for computing these effects 
(c.f. Hartmann and Boulet, \cite{HartmannBoulet91}), but these methods do not yet provide the 
accuracy or numerical efficiency required for routine use. LBLABC currently 
employs a simple, semi-empirical algorithm to correct for the effects of line
mixing.  For strongly mixed Q-branch lines, we use the first order correction 
described by Rosenkranz (\cite{rosenkranz88}). With the line mixing
 correction and the use of a
sub-Lorentzian Chi-factor (c.f. Fig 2 of Meadows and Crisp
 \cite{MeadowsCrisp96}), these
methods provide adequate accuracy for modeling the CO$_2$ absorption at pressures 
and temperatures as high as those seen in the deep atmosphere of Venus
(90-bars), and should provide the accuracy needed for the calculations proposed here.

These products, along with the atmospheric properties and constituent
mixing ratios were used as input into our line-by-line radiative
transfer model, SMART\_SPECTRA, which was developed by the Virtual Planetary 
Laboratory for use with generalized terrestrial planet climate models, and is 
an updated version of the Spectral Mapping Atmospheric Radiative Transfer 
model (SMART; Meadows \& Crisp \cite{MeadowsCrisp96}; Crisp \cite{crisp97})
developed by D. Crisp. 
For this application, SMART\_SPECTRA was used to calculate a high resolution 
spectrum of each planet for a solar zenith angle of 60$^\circ$, which
approximates the average illumination observed in a planetary
disk-average.  All spectra were generated over an ocean surface,
which provides a neutral background for understanding the atmospheric
changes in the models. The ocean surface, however, is extremely dark;
thus, the reflectivities shown here will typically be much lower 
than would be expected for the present Earth disk-average, which
would also include the higher-reflectivity contributions from
continents and clouds. For the visible and near-infrared spectra
shown here, the albedo is computed as the ratio of the integrated
upward flux over the hemisphere, divided by $\pi$ steradians, to
approximate the mean radiance seen in the disk-average, and then
divided by the solar insolation at the top of the atmosphere. For the
mid-infrared spectra, the radiance shown is the integrated upward
flux over the hemisphere, divided by $\pi$ steradians.

\section{Results}
\subsection{Early Earth standard atmosphere}
\begin{figure}
   \resizebox{\hsize}{!}{\includegraphics{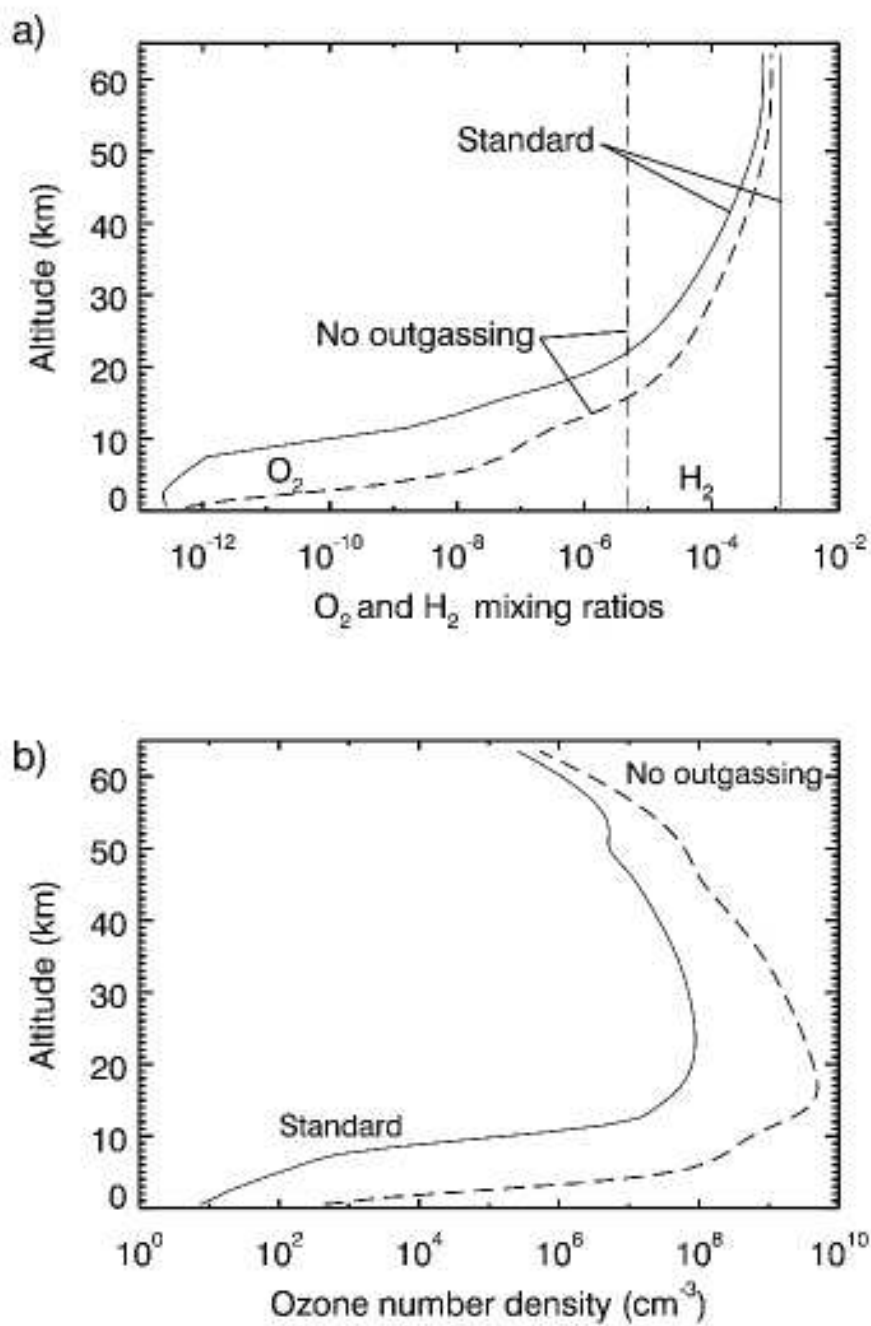}}
      \caption{O$_2$ and H$_2$ mixing ratios (a) and O$_3$ number densities (b) for the standard and 
      no outgassing cases.}
         \label{sunoutgasresults}
   \end{figure}

 The vertical profiles of O$_2$ and O$_3$ obtained in our ``standard'' atmosphere (0.2 bars CO$_2$ 
planet around the Sun) are presented in Fig.~\ref{sunoutgasresults}, and the associated columns depths 
are summarized in Table~\ref{concentrations}.
\begin{table*}
   \caption{Hydrogen budget for planets around the Sun and EK Dra (high UV) with 0.2 bars of CO$_2$.}
    \label{H2budget}
\centering                          
\begin{tabular}{l l c c c}        
\hline\hline
Chemical & Redox &\multicolumn{3}{c}{H$_2$ budget contribution$^{\mathrm{a}}$} \\
\cline{3-5}
species &coefficient & Standard & No outgassing & High UV  \\
\hline 
{\it Reduced species} \\
H         & 0.5  &  $4.88\times 10^{5}$ &  $8.80\times 10^{6}$ & $2.47\times 10^{5}$ \\
CO        & 1    &  $3.58\times 10^{7}$ &  $3.02\times 10^{5}$ & $2.04\times 10^{7}$ \\
HCO       & 1.5  &  $1.42\times 10^{8}$ &  $4.45\times 10^{7}$ & $2.18\times 10^{8}$ \\
H$_2$CO   & 2    &  $3.68\times 10^{9}$ &  $3.34\times 10^{8}$ & $9.16\times 10^{9}$ \\
H$_2$S    & 3    &  $2.16\times 10^{8}$ &  $3.84\times 10^{5}$ & $1.79\times 10^{8}$ \\
HS        & 2.5  &  $8.40\times 10^{7}$ &  $5.97\times 10^{5}$ & $6.40\times 10^{7}$ \\
HSO       & 1.5  &  $7.36\times 10^{7}$ &  $3.49\times 10^{5}$ & $4.55\times 10^{7}$ \\
NH$_2$    & 1    &  $8.02\times 10^{6}$ &  $6.75\times 10^{2}$ & $9.04\times 10^{6}$ \\
N$_2$H$_3$& 1.5  &  $1.24\times 10^{7}$ &  $8.10\times 10^{-1}$ & $1.47\times 10^{7}$ \\

N$_2$H$_4$& 2    &  $1.50\times 10^{8}$ &  $5.55\times 10^{-1}$ & $1.23\times 10^{8}$ \\
$\Phi_{\mathrm{rain}}(\mathrm{Red})$& & $4.4\times 10^{9}$ &  $3.89\times 10^{8}$ & $9.83\times 10^{9}$\\

{\it Oxidized species} \\
O              & $-1$   & $2.00\times 10^5$ & $5.91\times 10^6$ & $6.83\times 10^5$\\
HO$_2$         & $-1.5$ & $6.56\times 10^6$ & $6.73\times 10^6$ & $6.97\times 10^6$\\
H$_2$O$_2$     & $-1$   & $7.28\times 10^5$ & $2.68\times 10^7$ & $1.13\times 10^6$\\
HNO            & $-0.5$ & $2.63\times 10^7$ & $2.96\times 10^8$ & $3.47\times 10^7$\\
H$_2$SO$_4$    & $-1$   & $8.64\times 10^5$ & $6.97\times 10^6$ & $2.21\times 10^6$\\
SO$_4$ aerosol & $-1$   & $8.47\times 10^7$ & $1.88\times 10^8$ & $1.55\times 10^8$\\
$\Phi_{\mathrm{rain}}(\mathrm{Ox})$& & $1.19\times 10^8$ & $5.31\times 10^8$ & $2.01\times 10^8$ \\        

{\it Volcanism contribution}\\
H$_2$     & 1    & $2.00\times 10^{10}$ & 0.0                &   $2.00\times10^{10}$ \\
NH$_3$    & 1.5  & $3.87\times 10^{9}$ & $5.27\times 10^{6}$ &  $4.21\times 10^{9}$\\
CH$_4$    & 4    & $1.80\times 10^{10}$ & 0.0                & $1.80\times 10^{10}$ \\
$\Phi_{\mathrm{volc}}(\mathrm{H}_2)$& & $4.19\times 10^{10}$ & $5.27\times 10^{6}$ & $4.22\times 10^{10}$ \\
 & & & & \\
$\Phi_{\mathrm{esc}}(\mathrm{H}_2)$ & & $3.76\times 10^{10}$ & $1.47\times 10^8$ & $3.26\times 10^{10}$\\

H$_2$ balance                       & & $3.67\times 10^7$   & $6.02\times 10^3$& $3.53\times 10^{5}$\\
\hline 
\end{tabular}
\begin{list}{}{}
\item[$^{\mathrm{a}}$] In molecules cm$^{-2}$ cm$^{-1}$.
\end{list}
\end{table*}

\begin{table*}
   \caption{O$_3$ and O$_2$ column depths (cm$^{-2}$) and H$_2$ and CH$_4$ volume mixing 
            ratios for high CO$_2$ terrestrial atmospheres. 
            Present Earth values are shown for comparison.}
    \label{concentrations}
\centering                          
\begin{tabular}{l c c c c}        
\hline\hline
Case & Ozone & Oxygen (O$_2$) & Hydrogen & Methane \\
(bars of CO$_2$) & column depth$^{\mathrm{a}}$ & column depth & mixing ratio & mixing ratio \\
\hline 
{\it Present UV} \\
Present Earth & $8.61 \times 10^{18}$ & $4.65 \times 10^{24}$ & $5.5 \times 10^{-7}$ & $1.6 \times 10^{-6}$ \\ 
0.2 standard  & $1.87 \times 10^{14}$ & $1.24 \times 10^{19}$ & $1.25 \times 10^{-3}$ & $1.38 \times 10^{-4}$ \\
0.2 no outgassing & $7.51 \times 10^{15}$ &$4.94 \times 10^{19}$ & $4.85 \times 10^{-6}$ & $9.4 \times 10^{-12}$ \\
{\it Sensitivity tests$^{\mathrm{b}}$}\\
{\it No outgassing case modified}\\
O$_2$ \& O$_3$ third body reactions& $1.09 \times 10^{16}$ & $4.51 \times 10^{19}$ & $5.2 \times 10^{-6}$ & $9.2 \times 10^{-12}$ \\
H$_2$O absorption eliminated& $1.65 \times 10^{16}$ & $8.34 \times 10^{19}$ & $4.0 \times 10^{-8}$ & $2.1 \times 10^{-16}$ \\
$k_{\mathrm {edd}}$ increased & $6.84 \times 10^{15}$ & $2.78 \times 10^{19}$ & $3.8 \times 10^{-6}$ & $1.0 \times 10^{-11}$ \\
Stratospheric temperature 200 K&$3.66 \times 10^{15}$ & $4.38 \times 10^{19}$ & $5.8 \times 10^{-6}$ & $7.4 \times 10^{-12}$ \\
Stratospheric temperature 160 K&$1.05 \times 10^{16}$ & $4.64 \times 10^{19}$ & $4.7 \times 10^{-6}$ & $8.6 \times 10^{-12}$ \\
{\it Outgassing case modified}\\
Surface temperature = 240 K&$3.45 \times 10^{13}$ & $2.24 \times 10^{18}$ & $9.4 \times 10^{-4}$ & $2.7 \times 10^{-4}$ \\
{\it Higher UV} \\
0.02 & $3.53 \times 10^{14}$ & $6.62 \times 10^{18}$& $8.29 \times 10^{-4}$ & $1.66 \times 10^{-5}$\\
0.2  & $6.79 \times 10^{14}$ & $3.63 \times 10^{19}$& $1.09 \times 10^{-3}$ & $4.07 \times 10^{-5}$ \\
2.0  & $5.13 \times 10^{14}$ & $4.24 \times 10^{19}$& $1.17 \times 10^{-3}$ & $8.75 \times 10^{-5}$  \\
\hline 
\end{tabular}
\begin{list}{}{}
\item[$^{\mathrm{a}}$] Value for present Earth from McClatchey et al. (\cite{mclatcheyetal71}).
\item[$^{\mathrm{b}}$] See text for detailed descriptions.
\end{list}
\end{table*}
To illustrate the role of volcanic outgassing of reduced gases in determining O$_2$ and O$_3$ concentrations,
we performed a simulation with H$_2$ and CH$_4$ outgassing turned off.
As expected, more O$_2$ and O$_3$ were formed in the no-outgassing case. The calculated 
increases relative to the standard case were: $\sim$4 times for O$_2$ and a factor of $\sim40$ for O$_3$ 
(Table~\ref{concentrations}). The latter increase is, of course, large in a relative sense; however, 
the absolute ozone column density for the no-outgassing case is still $< 0.1$\%
 of the mean ozone column density on present Earth. The reason why the calculated increases in O$_2$ and 
O$_3$ were so modest is that H$_2$ did not disappear from the atmosphere even though volcanic outgassing was
turned off (Fig.~\ref{sunoutgasresults}a).
The H$_2$ remaining in this no-outgassing case is produced by atmospheric photochemistry, followed by 
rainout of oxidized species from the model atmosphere (Table \ref{H2budget}).

A set of sensitivity tests was performed in order to quantify some of the uncertainties in our model. 
Most of these test were done for the ``no outgassing'' case, which maximizes the amount of abiotic O$_2$ and O$_3$. 
Five specific modifications were tested in the model, the results of which are listed in Table \ref{concentrations}.
First, we doubled the third body reaction rates for O$_3$ and O$_2$ formation to simulate the possible effects 
of having CO$_2$ as a third body. This caused a modest ($\sim30$\%) increase in the column depth of O$_3$ and 
actually decreased the column depth of O$_2$.
Second, we eliminated the water absorption for wavelengths greater than 195 nm where the absorption coefficients
have been extrapolated from measurements at shorter wavelengths. This caused a 2-fold increase in both O$_2$ and
O$_3$, presumably due to the slower recombination of O$_2$ and H$_2$ when the abundance of H and OH radicals (from
H$_2$O photolysis) is reduced.
Third, the minimum of the eddy diffusion coefficient ($k_{\mathrm {edd}}$) was
increased by 3 (Fig. \ref{eddy}) in order to 
simulate the effect of no having an
inversion layer due to the lack of O$_3$ heating in the atmosphere as explained by Kasting et al. (\cite{kastingetal79}).
This decreased O$_2$ by about 40\% as a consequence of faster transport of O$_2$ from the upper stratosphere,
where it is produced from CO$_2$ photolysis, to the troposphere, where it is consumed by reaction with H$_2$.
We then raised and lowered the assumed stratospheric temperature by 20 degrees. Lowering the stratospheric 
temperature raised the O$_3$ column depth by $\sim 30$\%, while raising it lowered the column depth by a factor of 2.
Both effects are caused by the increased rate of O$_3$ destruction at high temperatures.

We also performed one additional sensitivity test in which the planet's surface temperature 
was reduced to 240 K. This is similar to the ``Snowball Earth'' simulation of Liang et al. 
(\cite{liangetal}). In this case, we did {\it not} turn off volcanic 
outgassing of reduced gases, because as noted in the Introduction, this is one of the known ``false positive''
that could lead to arbitrarily high levels of atmospheric O$_2$. Liang et al. 
did not predict high O$_2$ levels, but that is because they assumed that photochemically produced oxidants, 
specifically H$_2$O$_2$, would accumulate in the ice. Clearly, that could not continue indefinitely, 
and so such a simulation does not represent a true steady state. In our simulation, with reduced 
volcanic gases assumed to be present, the predicted concentrations of O$_2$ and O$_3$ were both low 
(Table \ref{concentrations}).
However, abiotic CH$_4$ concentrations increased to $\sim 2.7 \times 10^{-3}$ in this simulation, as a consequence 
of the reduced concentrations of tropospheric H$_2$O and correspondingly low number densities of OH radicals. 
Hence, this simulation reinforces the thought that one should be cautious about assuming that CH$_4$ 
by itself is a bioindicator, as there clearly are situations in which abiotic CH$_4$ might become abundant.

Although the sensitivity tests performed here by no means exhaust the possible sources of uncertainty in the 
photochemical model, they show that most such uncertainties lead to only modest changes in the results. Even a
factor of 4-5 uncertainty in the predicted O$_2$ and O$_3$ concentrations is not signficant if the resulting
column depths remain well below the level of detection. In general, the results are more sensitive to the 
boundary conditions imposed on the model, e.g., outgassing and escape of hydrogen, than they are to the details
of the model photochemistry.

\begin{figure}
   \resizebox{\hsize}{!}{\includegraphics{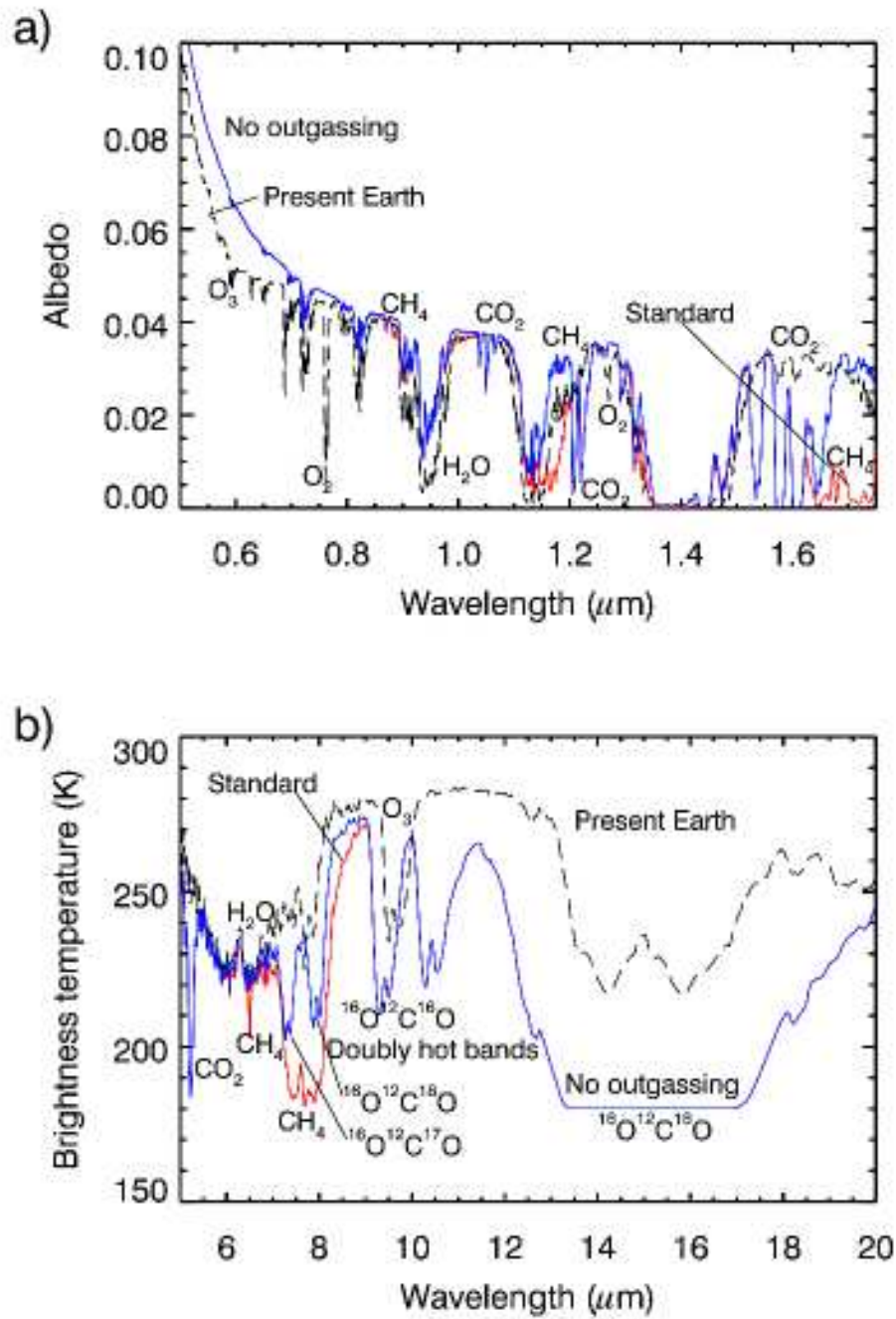}}
      \caption{Spectra of planets with 0.2 bars of CO$_2$ around the Sun. Standard case (red) and
               no outgassing case (dark blue).  Present 
               Earth spectrum is plotted for comparison (dashed line).}
         \label{spec_nooutgas}
   \end{figure}

We also generated spectra at visible and mid-infrared wavelengths for the 0.2 bar CO$_2$ standard early Earth
 case, both with and without outgassing, and we compared these to the spectrum of modern day Earth 
(Fig. \ref{spec_nooutgas}).    
In the visible-near-IR spectra (Fig. \ref{spec_nooutgas}a), the principal differences between the two early Earth cases can be seen in the 0.86-0.89 $\mu$m, 0.99-1.03 $\mu$m, 1.15-1.20$\mu$m and 1.62-1.70 $\mu$m wavelength ranges, where the early Earth with 
outgassing shows more pronounced absorption from methane, as is expected given
 that the no-outgassing case 
has a $\sim10^7$ times smaller concentration of this compound (Table~\ref{concentrations}). 
For O$_2$, although the no-outgassing case resulted in a 4-fold enhancement in the column depth, 
the total amount of O$_2$ was still sufficiently small that there is no sign of any absorption 
from the O$_2$ A-band at 0.76 $\mu$m in either case.  Comparison with the modern Earth spectrum 
also shows a significant and near-identical 
reduction in O$_3$ absorption (0.5-0.7 $\mu$m) for both early Earth cases.   Again, the higher O$_3$ column depth 
for the no-outgassing case was still too small to produce a discernible difference in the spectrum. The early 
Earth spectra also show strong CO$_2$ absorption in a triplet of bands centered around 1.05 $\mu$m 
and a doublet of bands from 1.2-1.25 $\mu$m, features that are absent from the modern Earth spectrum.  
The strongest CO$_2$ absorption however is seen from 1.52-1.65 $\mu$m, and is present in the modern Earth spectrum, 
although at much reduced strength.
 
In the mid-infrared (Fig. \ref{spec_nooutgas}b), again, the differences in the two early Earth spectra are dominated by methane absorption 
near 7.7 $\mu$m, which is much more pronounced for the outgassing case.  Any difference in the ozone band 
at 9.6 $\mu$m, however, is masked by the CO$_2$ doubly-hot bands at 9.4 $\mu$m, as noted by 
Selsis et al. \cite{selsisetal02}.   
In comparison to the modern Earth spectrum, the early Earth MIR spectrum is dominated by the absorption of 
CO$_2$ at 5.3, 7.4, 8.0, 9.4, 10.5 and 15$\mu$m.    These bands are associated with different isotopes of CO$_2$.  
The $\nu_2$ fundamental of the dominant CO$_2$ isotope (\element[][16]{O}\element[][12]{C}\element[][16]{O}) 
produces the strongest  feature at 15$\mu$m, which has nearly doubled in apparent width for the 0.2 bar CO$_2$ 
case, when compared to Modern Earth.       
The bands at 9.4 and 10.4 $\mu$m are ``doubly hot bands'' of the dominant isotope, transitions from 
upper states to the 2nd excited state, rather than the ground state.  However, the bands near 7.4 and 
8$\mu$m are not hot bands, but rather $\nu_1$ fundamentals for the asymmetric CO$_2$ isotopes 
(\element[][16]{O}\element[][12]{C}\element[][17]{O} and \element[][16]{O}\element[][12]{C}\element[][18]{O}).  
This band is forbidden by a Fermi resonance in all symmetric CO$_2$ isotopes, which includes the dominant 
\element[][16]{O}\element[][12]{C}\element[][16]{O} molecule.  
The flat base of the 15$\mu$m CO$_2$ $\nu_{2}$ absorption in the early Earth case is produced by 
the isothermal stratosphere in our model, which will also serve to enhance the apparent molecular 
absorption in the spectrum. Should the planet's actual vertical temperature profile deviate from 
this assumption, then the observed features in the MIR may be less pronounced, even given the same 
atmospheric composition.   Collisionally-induced absorption was also included in the model used 
to generate these spectra, but it shows relatively little effect on the spectrum throughout this 
wavelength range at these CO$_2$ partial pressures.

\subsection{High CO$_2$ planetary atmospheres under high UV radiation}
\begin{figure}
   \resizebox{\hsize}{!}{\includegraphics{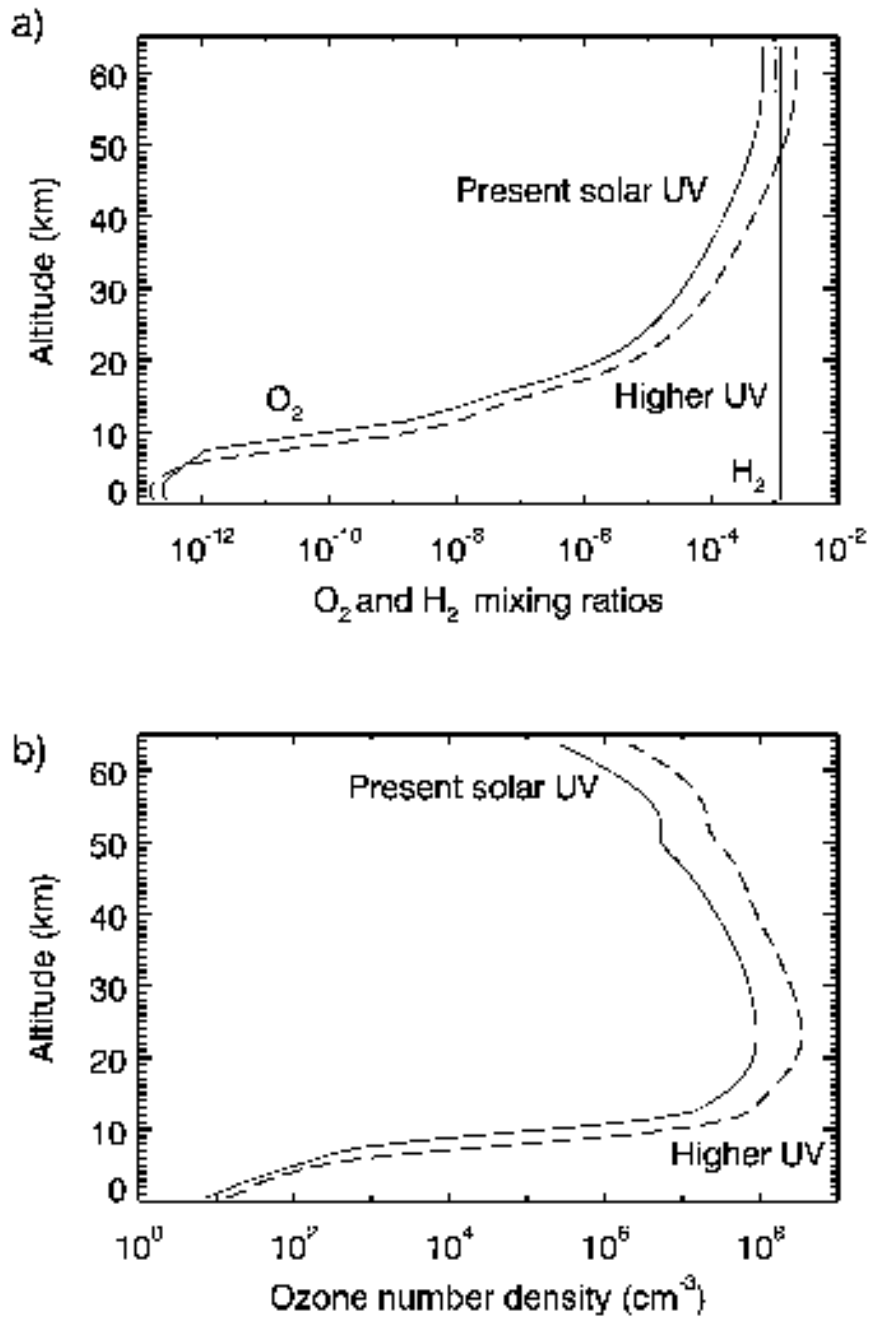}}
      \caption{O$_2$ mixing ratios (a) and O$_3$ number densities (b) for atmospheres with 
              0.2 bars of CO$_2$ under present UV solar radiation and high UV flux from a solar-like young star.}
         \label{lowhighUVresults}
   \end{figure}

\begin{figure}
   \resizebox{\hsize}{!}{\includegraphics{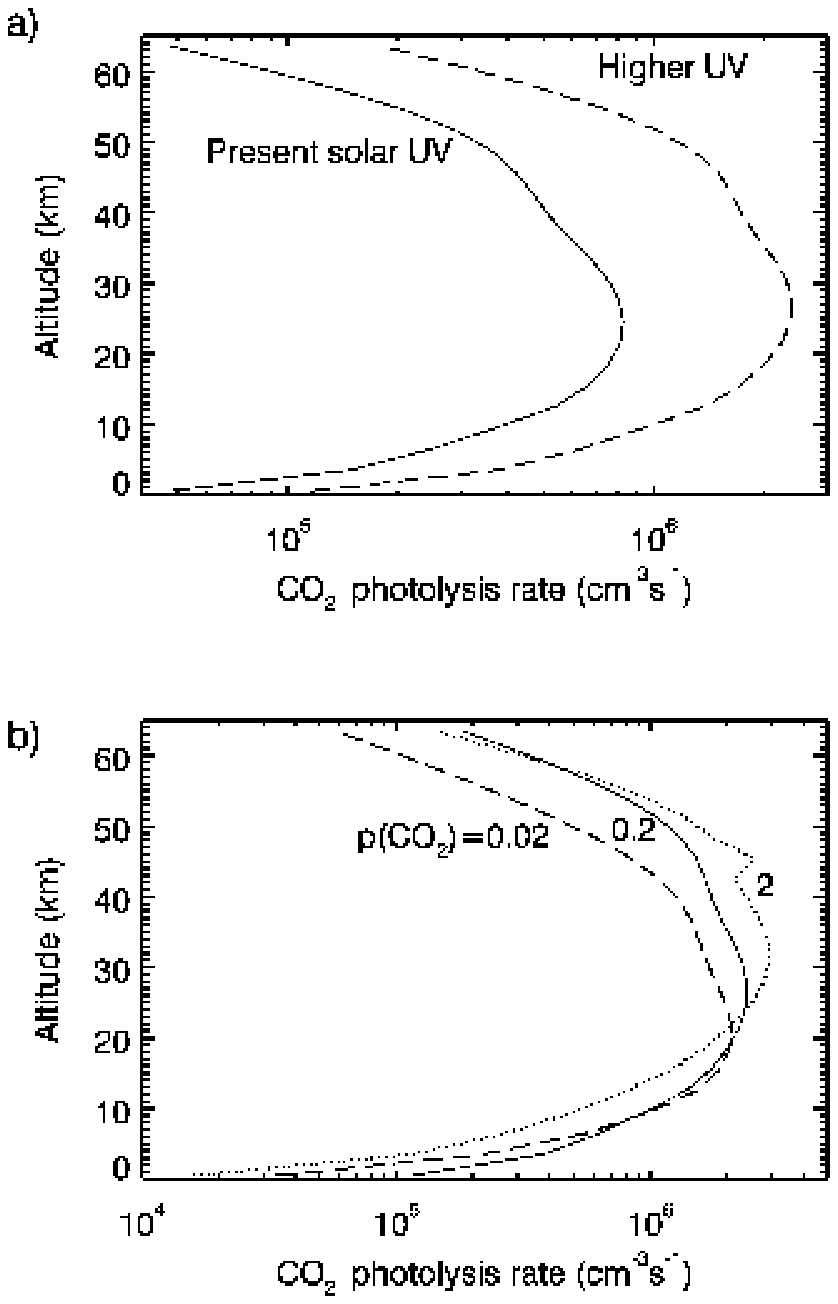}}
      \caption{CO$_2$ photolysis rates for
               {\it a)} atmospheres with 0.2 bar of CO$_2$ under present UV solar radiation
                  and high UV flux from a solar-like young star, and
               {\it b)} atmospheres under high UV radiation with different amounts of CO$_2$.}
         \label{photorates}
   \end{figure}

In this experiment, we simulated three planetary atmospheres, each with 0.02, 0.2 and 2 bars of CO$_2$, 
with high incoming UV radiation. The
purpose was to explore the effects of high UV insolation on abiotically produced O$_2$ and O$_3$ in high-CO$_2$ atmospheres. 
For the 0.2 bar-CO$_2$ atmosphere under high UV, 
the column depths of O$_2$ and O$_3$ were $\sim$3 times that of the comparable standard 
case with solar insolation (Tables \ref{concentrations} Fig.~\ref{lowhighUVresults}).
These concentrations, however, are still very low compared 
with the present Earth's ozone and oxygen column depths. H$_2$ amounts in the standard and high UV atmospheres were 
very similar (Fig.~\ref{lowhighUVresults}a), but larger photolysis rates of CO$_2$, resulted in more
O$_2$ (and hence, O$_3$) in the planet's stratosphere (see Fig.~\ref{photorates}a).

The ozone column depth changes very little when the CO$_2$ rises from 0.02 bars to 0.2 bars, given the same amount of 
UV radiation received at the top of the atmosphere. But, when the CO$_2$ is increased to 2 bars in a high UV environment, the ozone column 
depth decreases to 0.76 times that in the planet with 0.2 bar CO$_2$ around EK Dra.  
The likely reason is that
CO$_2$ photolysis occurred higher in the stratosphere, where densities are low and where 3-body formation of
O$_2$ and O$_3$ is inhibited (Fig.~\ref{photorates}). In any case, the calculated changes in O$_2$ and O$_3$ 
are too small to produce an appreciable effect on the spectra. By contrast, the result 
of going from the low-UV to the high-UV environment, keeping the same concentration of CO$_2$ in 
the planetary atmosphere, is to increase O$_2$ and O$_3$ column depths by 
a factor of $\sim$3  (Table~\ref{concentrations}).

   \begin{figure}
   \resizebox{\hsize}{!}{\includegraphics{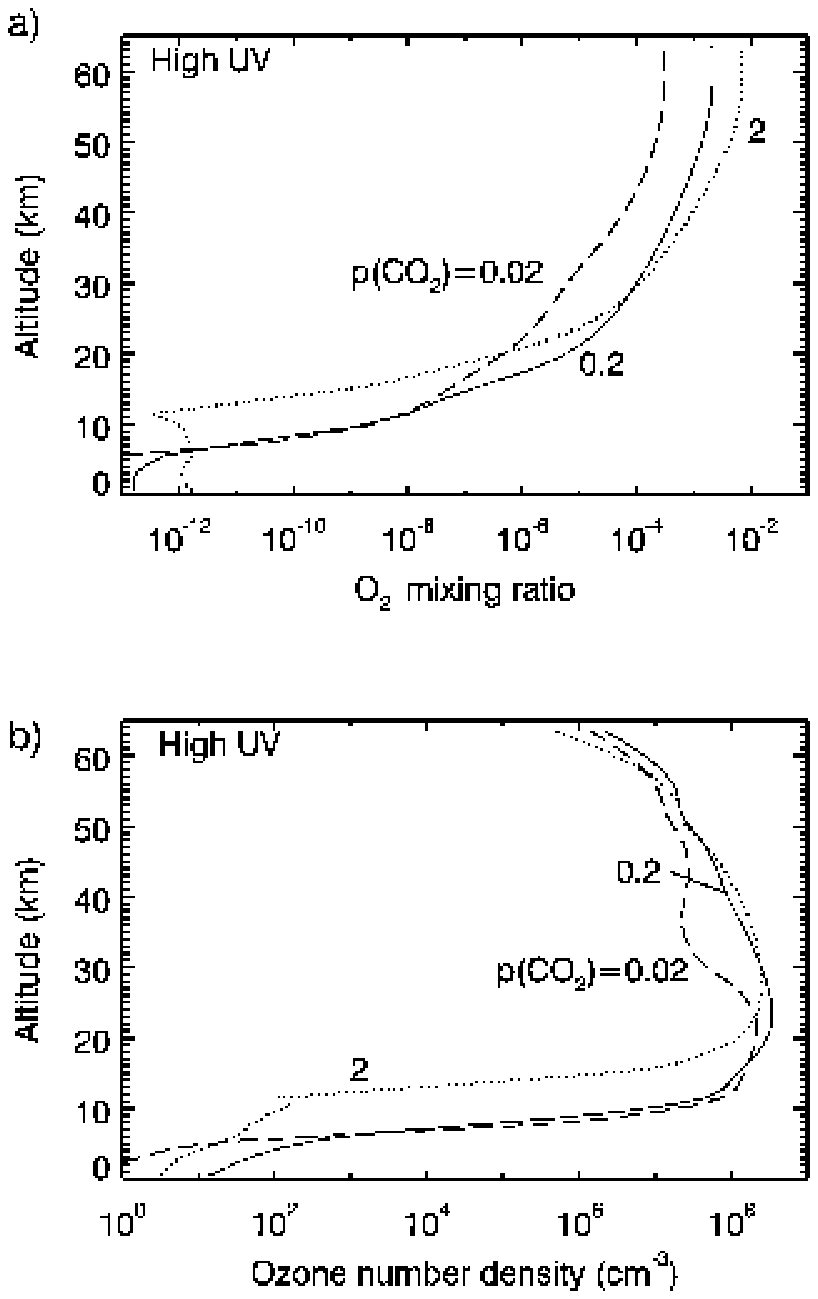}}
      \caption{O$_2$ mixing ratios (a) and O$_3$ number densities (b) for atmospheres with 
              different amounts of CO$_2$ under high UV flux from a solar-like young star.}
         \label{highUVresults}
   \end{figure}
Because the planet with the atmosphere containing 2-bars of CO$_2$ had a surface T (312K) 
significantly higher than our 0.02 and 0.2 bar cases (278K), we also simulated a planet around 
EK Dra with 2 bars of CO$_2$, but using the colder surface temperature of 278K (Fig. \ref{profiles}).   
This case was done to minimize the effect of the temperature on the ozone and oxygen destruction 
and to model an atmosphere more similar to early Mars. The O$_3$ and O$_2$ column depths for 
this case were less than those obtained for the hotter 2 bar-CO$_2$ planet around EK Dra, 
being $3.7 \times 10^{14}$ cm$^{-2}$ and $2.7 \times 10^{19}$ cm$^{-2}$, respectively.  

\begin{figure}
   \resizebox{\hsize}{!}{\includegraphics{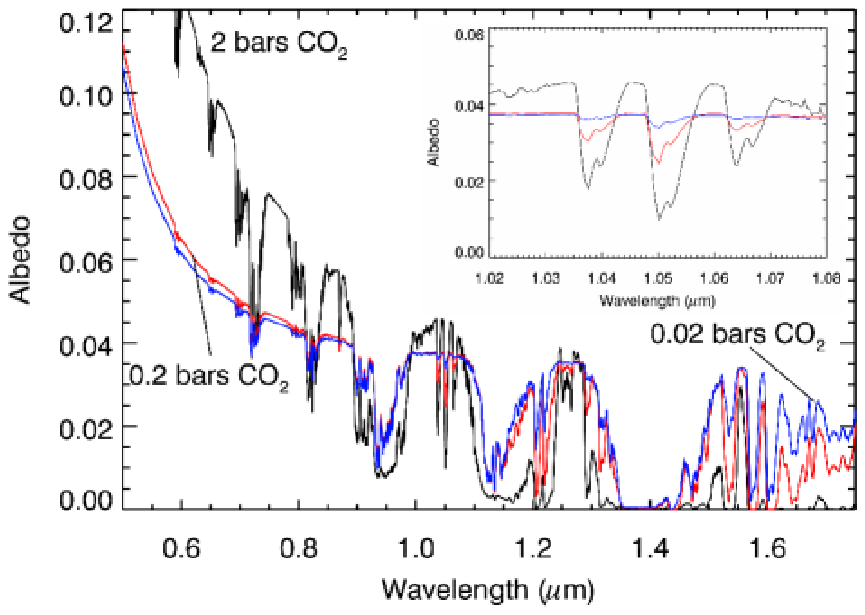}}
      \caption{Visible-NIR spectra of planets with 0.02 (dark blue), 0.2 (red) and 
               2 bars (black) of CO$_2$ around EK Dra.}
         \label{spec_3CO2_sol}
   \end{figure}

\begin{figure}
   \resizebox{\hsize}{!}{\includegraphics{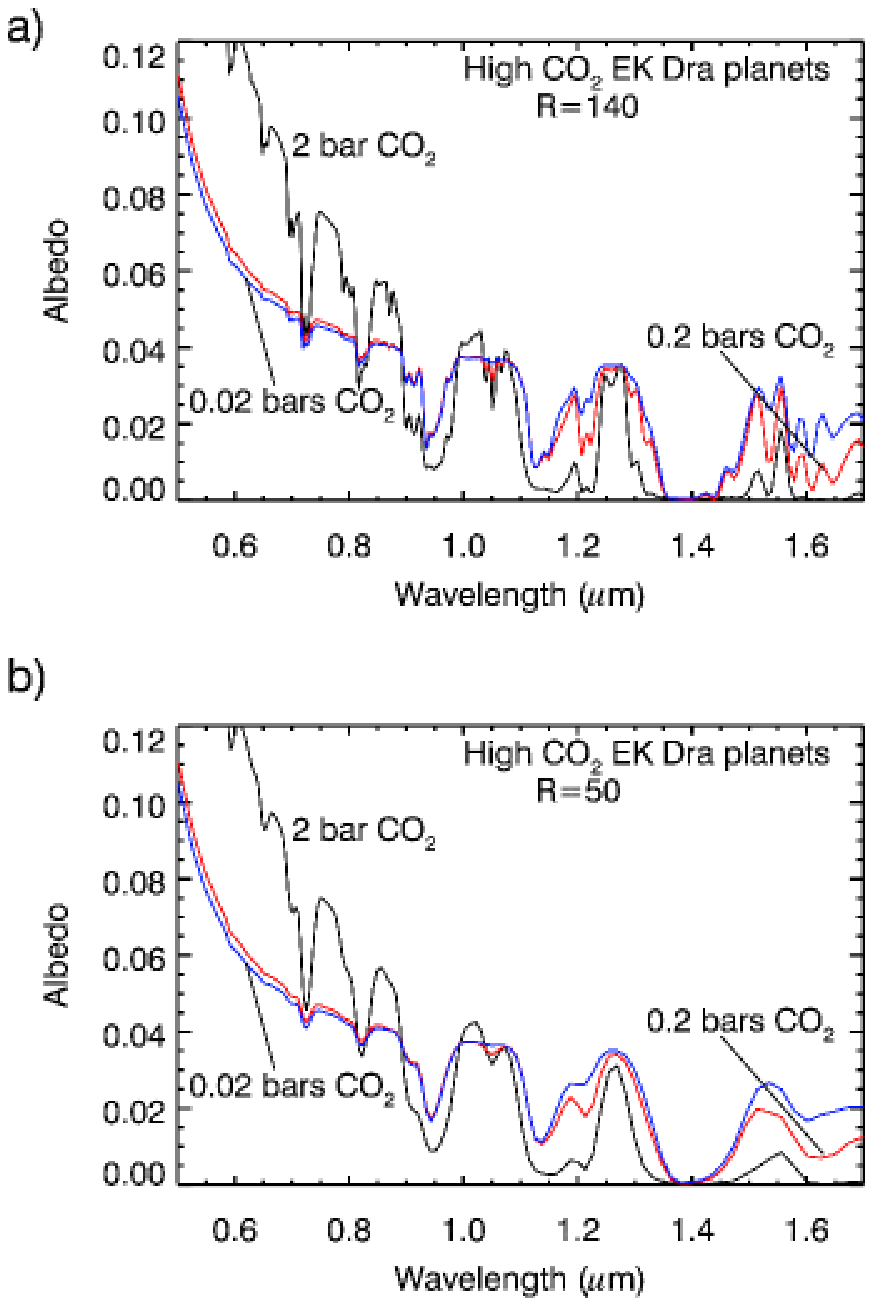}}
      \caption{Spectra for the 0.02, 0.2 and 2-bar CO$_2$ cases convolved with
               a triangular slit function and degraded to a constant wavenumber
               resolution equivalent to a)R $\sim$140, and b) R $\sim$50.}
         \label{spec_3CO2_R140-50}
   \end{figure}

\begin{figure}
   \resizebox{\hsize}{!}{\includegraphics{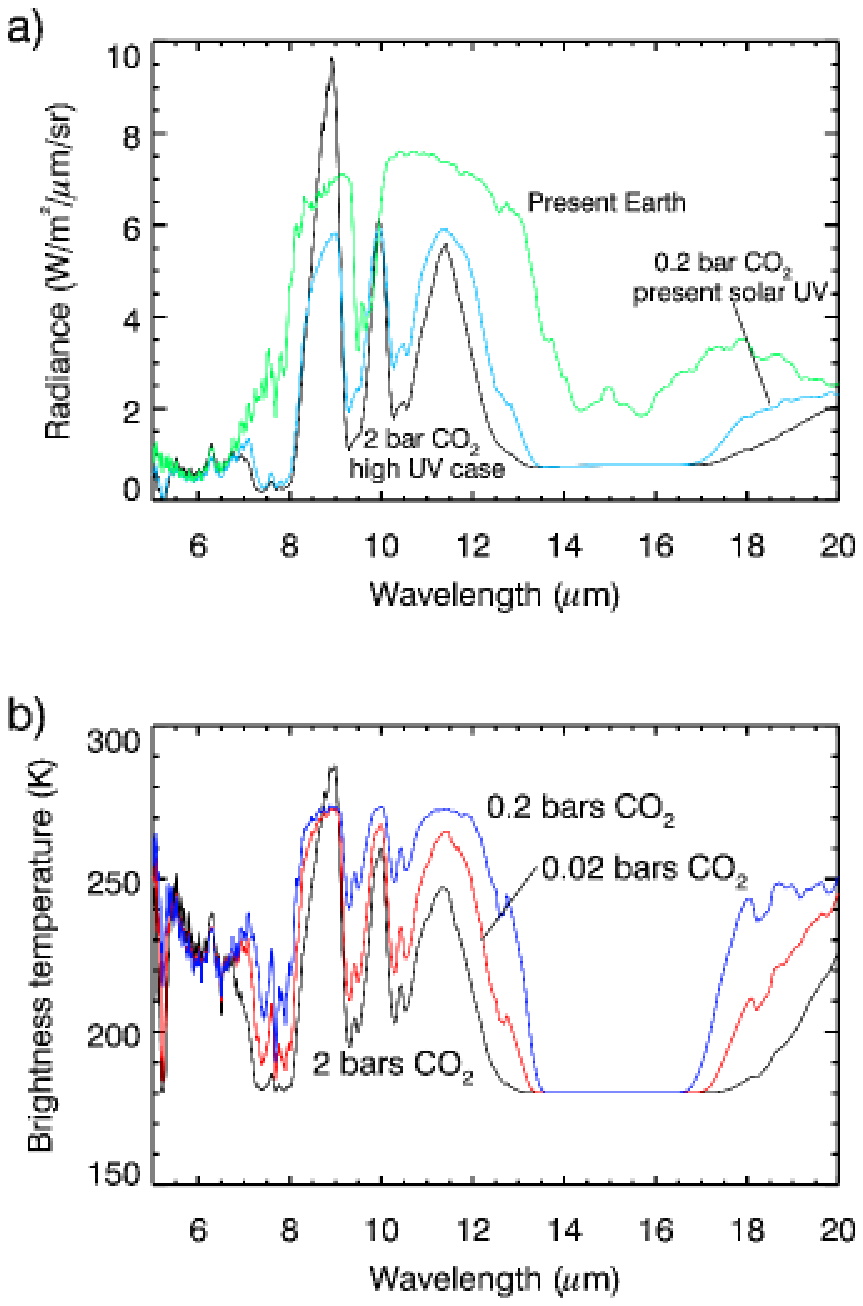}}
      \caption{a) MIR spectra for present Earth (green), our standard case 
(blue, 0.2 CO$_2$ bar, present solar UV) and 2 bars (black) 
of CO$_2$ around EK Dra. 
b) MIR spectra of planets with 0.02 (dark blue), 0.2 (red) and
2 bars (black) of CO$_2$ around EK Dra.}
         \label{spec_3CO2_ir}
   \end{figure}

The spectra for these three high-UV cases (Fig. \ref{spec_3CO2_sol} to  \ref{spec_3CO2_ir}) do show
differences from the standard model spectrum.  The most obvious difference in the visible spectrum is the change in the slope of the spectrum 
between 0.5-0.7$\mu$m. This slope can be affected by both O$_3$ absorption and Rayleigh scattering from atmospheric gases.   Since the total O$_3$ column depth for the 0.02, 0.2 and 2-bar CO$_2$ atmospheres is nearly identical, the difference in the reflectivity in this wavelength range is instead caused by enhanced Rayleigh scattering, and is not the result of differences in ozone absorption.   This scattering enhancement is due to either changes in total atmospheric pressure, or atmospheric composition  The more massive 2-bar CO$_2$ atmosphere has a surface pressure of 2.9 bars, compared with 1 bar for both the 0.02 and 0.2 bar CO$_2$ cases, and its large enhancement in the Rayleigh scattering slope between 0.5-0.7 $\mu$m is due primarily to its larger atmospheric pressure.   However, the 0.02 and 0.2 bar CO$_2$ atmosphere cases also show a small difference in their Rayleigh scattering slopes, even though they have identical atmospheric pressures.  This is due instead to the difference in the composition of their atmospheres, as CO$_2$ is a more efficient Rayleigh scatterer than nitrogen, and enhances the Rayleigh scattering slope for the 0.2 bar CO$_2$ case over the 0.02 bar case. 
In addition to the Rayleigh scattering slope, CO$_2$ absorption features are
also seen strongly in the spectrum and reveal spectral sensitivity to
atmospheric CO$_2$.  The triplet of absorption bands near 1.05$\mu$m
produces no appreciable absorption for the 0.02 bar case, but can be
distiguished as a triplet at high resolution (R $\sim140$ and above) for both
the 0.2 and 2-bar CO$_2$ cases (Fig. \ref{spec_3CO2_sol} - inset). At
resolutions of R $\sim$50 and greater, the triplet structure cannot be clearly
discerned, but the ability to distinguish between the 2-bar CO$_2$ and the
other two CO$_2$ concentration cases at 1.05$\mu$m is maintained (Fig. \ref{spec_3CO2_R140-50}),
assuming high S/N measurements. The CO$_2$ bands at 1.206 and 1.221$\mu$m become
progressively stronger as the atmospheric CO$_2$ abundance is increased and
can also be distinguished as two features at R $\sim 140$, but still retain
sensitivity to atmospheric CO$_2$ abundance at resolutions as low as R $\sim
50$.  A similar trend is shown for the stronger CO$_2$ bands near 1.6 $\mu$m,
however in this case, the 2-bar CO$_2$ atmosphere is so heavily absorbing that
the planetary spectrum becomes black throughout much of the 1.52-1.7$\mu$m
region, with the exception of a relatively transparent, narrow atmospheric
window at 1.55$\mu$m.
 
Finally, marked differences in the strengths of the water bands near 0.94$\mu$m, 1.14$\mu$m and 1.4$\mu$m are observed for the 2-bar CO$_2$ case.  This enhanced absorption is the result of increased H$_2$O abundances for the 2-bar CO$_2$ case, as compared to the two lower-CO$_2$ cases (Fig.\ref{profiles}), due to the higher surface temperature.  The model tropospheres assume a fixed distribution of relative humidity; hence, H$_2$O abundances increase exponentially with increasing surface temperature, following the corresponding increase in the saturation vapor pressure.
The water absorption features can be identified, and different concentrations
discriminated between, at resolutions as low as R $\sim 20$.  However the ability to
separate other species from the water vapor (such as the 1.2$\mu$m CO$_2$ absorption
feature) at R $\sim 20$ becomes extremely difficult, instead requiring resolutions
closer to R $\sim 50-70$ to do so, even for high S/N.

In the MIR, the spectra are dominated again by absorption from CO$_2$, and
these early Earth spectra are not at all similar to that of modern Earth
(Fig. \ref{spec_3CO2_ir}a). In particular, for the 2-bar CO$_2$ case, the
spectrum is heavily distorted by the CO$_2$ absorption, which narrows the
Earth's normal atmospheric window region between 8-13$\mu$m down to only
8-9$\mu$m.  The effects of collisionally-induced absorption are more
pronounced here for the 2-bar CO$_2$ case, and manifest themselves in this wavelength
range primarily as a broadening of the CO$_2$ isotope bands, enhancing the
wing absorption near 7.4 and 8.0 $\mu$m.  
While there is some theoretical justification for CO$_2$ pressure-induced
continuum absorption within the 10 $\mu$m and 15 $\mu$m bands, existing
measurements show that this absorption is apparently very weak compared to the
allowed transitions at these wavelengths.  We are not aware of any
quantitative estimates of its intensity.  For example, recent high-pressure
(200 atmosphere) measurements by Hartmann's group (c.f. Niro et al. \cite{niroetal2004}) show $\sim10\%$ residuals with respect to their
best line mixing model in simulations of absorption at pressures at 200
atmospheres. Some of these residuals might be associated with
collision-induced absorption, which is neglected in their model.  In any case,
if this absorption scales as $p^2$, it should be $10^4$ times weaker for a 2-bar
atmosphere, and would therefore have a negligible effect on the modeled spectrum within the context of the work reported here.

In all three cases, methane absorption at 7.7$\mu$m is greatly enhanced over
that of modern Earth, given that higher concentrations of this compound are
present in these low-O$_2$ cases (Table \ref{concentrations}). Looking at the MIR brightness temperatures for these three cases (Fig.\ref{spec_3CO2_ir}b), the atmospheric window in the 2-bar CO$_2$ case exhibits an enhanced brightness temperature of 287K when compared with the 272K brightness temperature for the other two cases.   This is due to the fact that the atmosphere is considerably hotter overall, but note that the surface temperature for the 2-bar CO$_2$ case is 317K, and for the other two, 278K.  In all cases, even though no clouds were present, it was not possible to retrieve the surface temperature of the planet by measuring the brightness temperature in the atmospheric window.   This is due to the fact that the atmospheric window is not completely transparent, but includes weak absorption from water vapor in an Earth-like atmosphere, and potentially other absorbers in a non-Earth-like atmosphere.  The water vapor, which is much enhanced in the 2-bar CO$_2$ case has effectively truncated the atmospheric column at a level higher and cooler than the surface.   For the 2-bar atmosphere the 30K discrepancy between the actual surface and the measured brightness temperatures corresponds to a sampled altitude of $\sim 4$km above the surface. 

Direct comparison between spectra generated for the 0.2 bar CO$_2$ planetary
atmospheres for high-UV (EK Dra) and low-UV (Sun) cases show very few
differences, except for changes in the strength of the methane absorption
features for both the vis-NIR and MIR.   This is due to the UV-induced removal
of CH$_4$ from the EK Dra planet, which reduces the overall CH$_4$ mixing
ratio by a factor of $\sim$3 for the high-UV case, as can be seen in Table \ref{concentrations}.   

\begin{figure}
   \resizebox{\hsize}{!}{\includegraphics{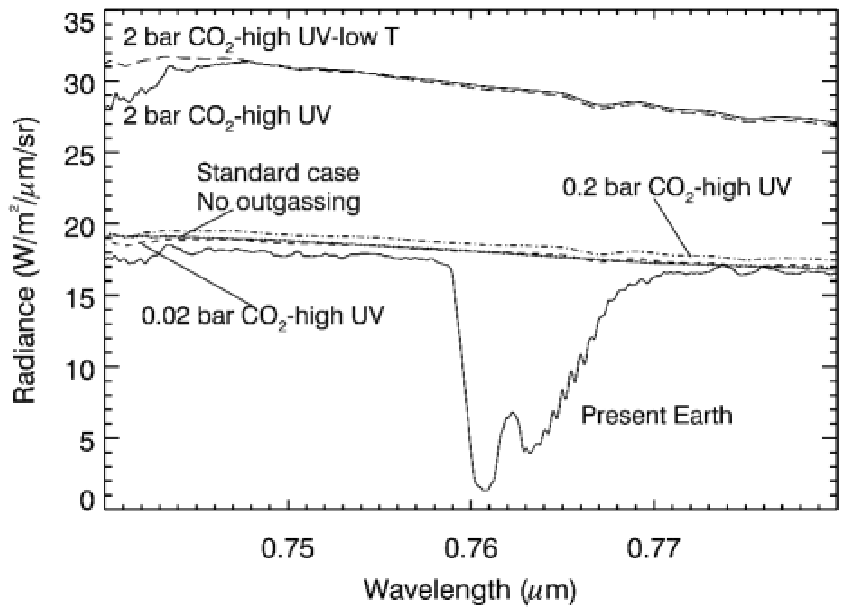}}
      \caption {O$_2$ A-band for all the simulated atmospheres.}
         \label{spec_O2band}
   \end{figure}

Finally, the modeled strength of the O$_2$ A-band feature in the visible, for
 all the planetary cases considered here (Fig.\ref{spec_O2band})
 demonstrates that the generation of a detectable false positive for oxygen is highly
 unlikely, even with very high spectroscopic resolution and S/N.

\section{Discussion}

Our results show that for a habitable planet, rainout of oxidized species onto a reduced surface will inhibit 
the generation of an abiotic ``false positive'' signal with strong O$_2$ or O$_3$ absorption.   As summarized in 
Figure \ref{spec_O2band}, this result is robust to both high stellar UV output and high atmospheric CO$_2$ abundances,
 as abiotic O$_2$ and O$_3$ are not visible for any of our cases, even in the extreme case of a planet with an
 atmosphere containing 2-bars of CO$_2$ and in orbit around a high-UV star.  We also explored the sensitivity of atmospheric 
constituents and their detectability to volcanic outgassing rates, incident UV flux, and atmospheric composition.  
 We found that the most significant spectral effects of either turning off volcanic outgassing, or increasing UV 
flux, was to noticeably decrease the observed methane absorption.  Increased stellar UV flux had little observable 
effect on the rest of the planetary spectrum, even though modest enhancements were seen in the atmospheric O$_3$ 
and O$_2$ column depths.  

Our results contradict previous models (Selsis et al. \cite{selsisetal02}) that predicted high abiotic abundances 
of O$_2$ and O$_3$ in planets with CO$_2$-rich atmospheres.  As explained earlier, these models failed to balance the H$_2$ 
budget because they ignored the rainout of soluble oxidized and reduced compounds. Including these terms--particularly
 the rainout of oxidants such as H$_2$O$_2$, H$_2$SO$_4$, and HNO, ensures that an abiotic atmosphere on an Earth-like
planet should contain substantial concentrations of H$_2$ and only minimal amounts of O$_2$ and O$_3$.

That said, there still may be planets within the habitable zones
  of their parent stars that do not have active hydrological cycle for which calculations similar 
to those of Selsis et al. may be valid and on which abiotic O$_2$ and O$_3$ could conceivably build up. For example, if
a planet in the habitable zone lacked significant liquid water, perhaps because it formed  with no appreciable water inventory, as has been modeled for terrestrials in planetary systems with eccentric Jovians
(Raymond et al. \cite{raymondetal04}), then our entire
analysis of the hydrogen budget would be invalid. H$_2$ would not be produced by reaction of photochemically produced
oxidants with reduced species in the crust and ocean, nor would it be likely to be outgassed from volcanoes. This 
would be compensated by a complete lack of hydrogen escape to space, thereby eliminating the net source of O$_2$ 
identified earlier in this paper. Some abiotic O$_2$ would presumably
  accumulate in the atmosphere due to photolysis of CO$_2$, although we have
  not attempted to quantify O$_2$ production for dessicated planets in this
  paper. However, abiotic production of O$_2$ on 
such a water-free planet could be relatively easy identified by the complete
  lack of H$_2$O bands in the planet's visible and MIR spectrum, which would
  indicate that it was unlikely to be an abode for Earth-like, water-and-carbon-based life.

Another exception to our analysis would be a frozen, ``Snowball Earth'' type planet within
the habitable zone of its star. Earth itself appears to have undergone global glaciation at least 3 times during its
history, most recently in the Neoproterozoic, at $\sim610$ Myr and 720 Myr
ago (Hoffman et al. \cite{hoffmanetal98}).  On such a frozen planet, as on the dry planet
described above, photochemically produced oxidants would not react with reduced materials in the planet's crust and 
ocean. The planet would thus be similar to the Mars-like planet mentioned in
the introduction for which atmospheric O$_2$ could, in theory, continue to
accumulate leading to a possible ``false positive'' 
for life.

Two things mitigate against this posing a significant problem for life detection by TPF or Darwin. One is that
Snowball Earth episodes on Earth are thought to be relatively short-lived, 10's of millions of years or less, because 
they are eventually terminated by buildup of volcanic CO$_2$ in the atmosphere (Hoffman et al.\cite{hoffmanetal98}). 10
Myr represents only about 0.2\% of the Earth's geologic history, and
the odds of finding an Earth-like planet in such a state in our Solar
neighborhood are relatively small. Second, it should be possible to
identify whether a planet is or not in a Snowball Earth-type phase, either by
looking for the presence of water ice or water vapor bands in their spectrum, or
by follow-up measurements of polarization to characterize the planetary surface and search for the presence of liquid water.

\begin{figure}
   \resizebox{\hsize}{!}{\includegraphics{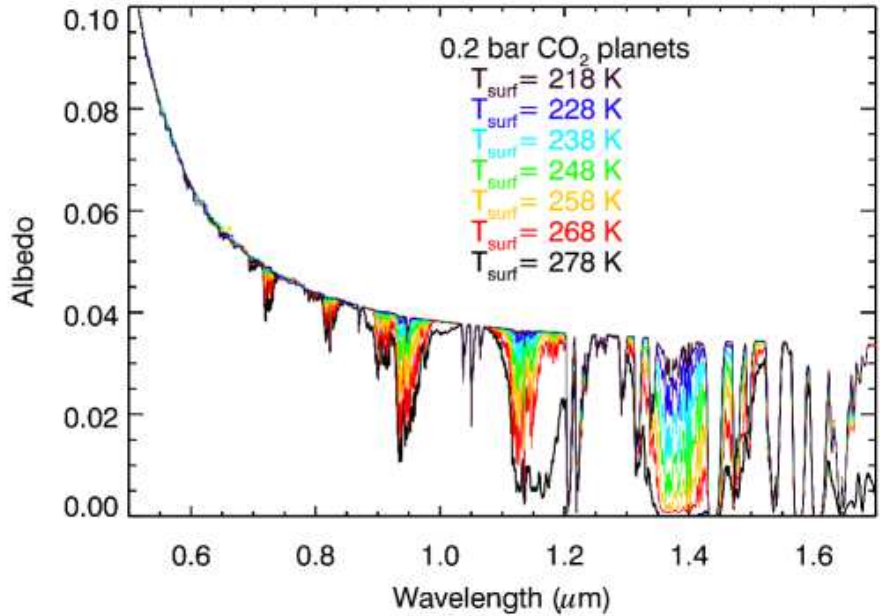}}
      \caption{Albedo of planets with 0.2 bars of CO$_2$ around the Sun with
        different surface temperatures.}
         \label{spec_lowT_sol}
   \end{figure}

\begin{figure}
   \resizebox{\hsize}{!}{\includegraphics{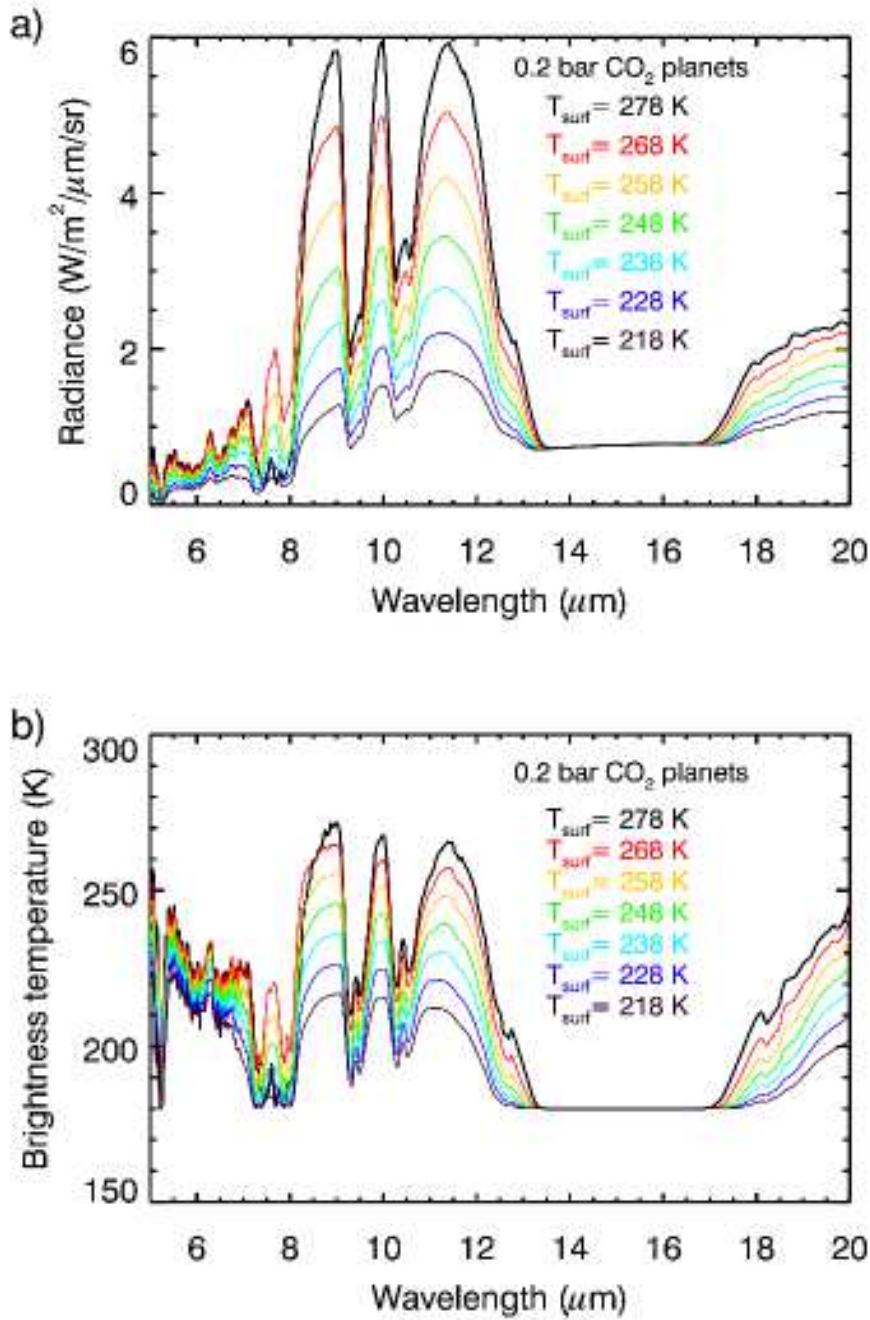}}
      \caption{MIR spectra of planets with 0.2 bars of CO$_2$ around the Sun with
        different surface temperatures.a) Radiance b) Brightness temperature}
         \label{spec_lowT_ir}
   \end{figure}

To quantify the effect of decreasing global surface temperature
  on the detectability of water vapor we  ran a series of models similar to
  our standard, 0.2 bar CO$_2$ case, but for successively lower surface
  temperatures. The corresponding  synthetic spectra generated from these
  environments are shown in Figs. \ref{spec_lowT_sol} and  \ref{spec_lowT_ir}.  Model calculations of Snowball Earth (Hyde et al. \cite{hydeetal2000})
show that the surface temperature drops precipitously from $\sim275$ K to 230 K when global glaciation sets in. For an
atmosphere with Earth's relative humidity distribution, Fig. \ref{spec_lowT_sol}
  shows that the water vapor bands in the visible and near-IR
become extremely weak as this transition occurs 
with a 238K surface temperature planet showing only $\sim$ 5\% of the
  absorption strength in the 1.13$\mu$m water band when compared to the
  feature seen for the 278K surface temperature planet.  However, with high
  S/N and a spectral resolution greater than R $\sim$50, the 1.13$\mu$m feature
  could be detected at temperatures greater than  238K.  It must be stressed
  that S/N and resolution are trade offs, as lower resolutions can require
  higher S/N to detect a given feature.  In the MIR (Fig. \ref{spec_lowT_ir}) the
  changes in the spectrum with decreasing surface temperature are seen as an
  interplay between lower water vapor amounts and the atmospheric
  temperature. The declining planetary atmospheric temperatures with surface
  temperature are seen in these cloud-free cases in the atmospheric window
  region between 8-13$\mu$m.  Although note that even in these cloud-free
  cases, the peak temperature detected in the atmospheric window at 9$\mu$m is
  not the surface temperature, being several degrees cooler due to the
  presence of water vapor, which results in atmospheric temperatures sampled
  at levels higher than the surface (Fig. \ref{spec_lowT_ir}b). If clouds were present, the inferred
  temperature would typically be even colder. The declining water vapor
  amounts are seen in the difference in the brightness temperature between the
  5.5-7.0$\mu$m water vapor region and the 8-13$\mu$m window region.  Correct
  interpretation of these spectra, however, would require generation of
  synthetic spectra that could be used to disentangle the combined effects of
  water and temperature on the spectrum.  Nonetheless, as was the case more
  straightforwardly for the visible-NIR spectrum, it may be possible to detect
  the significantly lower water vapor amount, given instrumentation that was
  sensitive to the 6.3$\mu$m water vapor band.  In this case, a resolution
  greater than R $\sim20$ in the MIR might be sufficient, with high S/N.  However,
  we add the caveat that the presence of clouds, either planet-wide, or
  covering a smaller fraction of the visible planetary disk, could skew the
  interpretation to colder temperatures and lower water abundances than are
  present at the planet's surface.  For these cases, searching for abundant
  water-soluble atmospheric gases, such as SO$_2$ might serve as a secondary
  indicator that we are observing a planet without a surface ocean, rather than
  one with clouds obscuration.  Multiple samples of the planet's atmosphere
  over time may also help to determine whether clouds are affecting the
  observed temperatures and water abundances.


Another way to test for the presence of liquid water on an exoplanet's surface is to measure the
polarization of light reflected from its surface. The specular ``glint'' produced by the smooth ocean surface
should be strongly polarized, at least at some phase angles (McCullough
\cite{mccullough07}; Stam et al. \cite{stametal06}; West et al. \cite{westetal05}; Williams \&
Gaidos \cite{WilliamsGaidos04}).
 Separation of different polarizations is
a feature of some suggested techniques for coronagraphic starlight suppression. Even if this measurement was not 
performed by an initial TPF-C mission, it could presumably be done by some follow-up mission as a way of verifying
the presence of surface liquid water (or at least of some surface liquid). A positive detection of either O$_2$ or
O$_3$ in an exoplanet's atmosphere would certainly warrant further investigation to be sure that the liquid water 
required for life was present at the planet's surface.

\subsection{Applicability to other planetary systems}

For these simulations, we used EK Dra, a young, UV-active star with a calculated age of only $10^8$ years.
  If we were considering only whether this star was likely to harbor habitable planets, then this would be a poor 
choice,  as planets around a star of this age would still be undergoing large, ocean-vaporizing impacts
 (Sleep et al. \cite{sleepetal}).  We would not anticipate seeing planet-wide biosignatures at this early and 
hostile stage, and so would also be unlikely to mistake a false positive as a biosignature.  Nonetheless,
 EK Dra is an excellent example for our simulations, as it most realistically represents the approximate 
upper limit on the UV radiation likely to be encountered by a planet around a Sun-like star.         

When the first studies of the UV environment for early Earth were published, the only known examples 
of young Sun-like stars were the T Tauri type stars. IUE observations showed that T Tauri UV fluxes were as much 
as $10^4$ times higher than the present solar UV radiation.  Based on those observations and stellar evolution 
models, those early studies concluded that the UV enhancement for a $10^8$ year old, Sun-like star, should be of 
the order of 10 times the present solar UV flux (Canuto et
al. \cite{canutoetal82}, Zanhle \& Walker \cite{zahnlewalker82}).    However,
we believe that this is an overestimate. EK Dra, at $10^8$ years,
and one of the most UV active stars known (Ribas et al.\cite{ribas}), does
exhibit fluxes 10 times that of the Sun at wavelengths shortward of 160nm.
However, when this flux is integrated over the wavelength range most relevant
to atmospheric chemistry ($\lambda < 200$ nm), it is only $\sim4.5$ times
larger than that of the Sun, on average.   More importantly for the
atmospheric chemistry, the continuum photon flux (rather than the energy flux)
at $\lambda < 200$ nm is just $\sim 3.6$ times higher in EK Dra when
compared to the solar spectrum (c.f. \ref{uvflux}) (Note that our continuum
calculations do not include the Ly$\alpha$ line which we calculated to be
$\sim 10$ times larger than for the present Sun, and which was included in all
our modeling simulations).   These considerations suggest that our results can
be interpreted as an upper limit on detectability for high-CO$_2$ atmospheres
around young, UV-active stars that are either Sun-like, or cooler than our
Sun.  If we consider stars with higher-UV flux than our Sun, e.g. the F stars,
it is also highly likely that our results are applicable.  An F2V star has an
integrated UV flux that is approximately 2.5 times that of EK Dra
(Segura et al. \cite{seguraetal03}). An examination of our Table 2 indicated that the increase in O$_2$ and O$_3$ column depth in a high-CO$_2$ atmosphere as a result of EK Dra's 4.5 times higher flux is relatively modest, being 3.6 and 2.9 times respectively, and this is for column depth values that are still 10$^3$ times lower than those likely to be detectable for a planet around an F2V star (see Segura et al. \cite{seguraetal03}, Table 1 and Fig 14).  It is therefore unlikely that the additional increase of 2.5 in the UV flux from an F star would result in detectable O$_2$ and O$_3$ for the high-CO$_2$ planets modeled here.  We therefore conclude that the results presented here are an upper limit for the majority of parent stars likely to be considered as TPF/Darwin targets.

\subsection{Effect of clouds on our conclusions}

\begin{figure}
   \resizebox{\hsize}{!}{\includegraphics{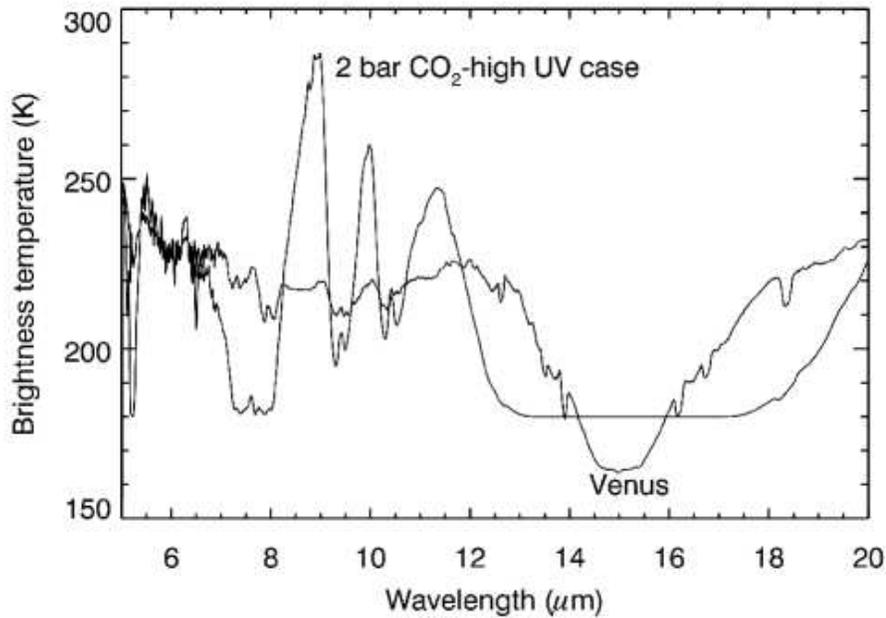}}
      \caption{Spectra of the 2 bars of CO$_2$ planet around EK Dra and Venus.}
         \label{spec_Venuscompar}
   \end{figure}

All the simulations shown here were run for clear-sky atmospheres, with no
appreciable hazes or cloud layers, and isothermal stratospheres.  These cases
shown here are therefore the most sensitive for detecting O$_2$ and O$_3$ in
our planetary spectrum, as they allow a longer pathlength through the
atmosphere that is not truncated by clouds.  Additionally, the contrast
between the hot surface and lower atmosphere, and the cool, isothermal
stratosphere provides for maximum detectability in the mid-infrared.
However, given that both the O$_3$ and O$_2$ concentrations peak above the
tropopause in all our model cases, and have significantly smaller abundances
below that, a cloud at the tropopause would not significantly change the net
column abundance of O$_2$ and O$_3$, and hence their detectability.   In the
case of the CO$_2$ however, which is evenly mixed through our atmospheres, a
cloud at the tropopause would significantly truncate the observed column, and
would result in much reduced absorption depths for CO$_2$ in the visible, and
would also dramatically reduce the CO$_2$ absorption in the mid-infrared, as
the cooler clouds tops would obscure the hotter surface and lower atmosphere
and provide a reduced temperature difference with the stratosphere. The
brightness temperatures of the observed emitting layer would also be reduced
to that of the cloud top temperatures, and would preclude our ability to
determine the planet's surface temperature.  This case can be demonstrated by
comparing the brightness temperatures for the 2-bar CO$_2$ atmosphere and a
modeled spectrum of the planet Venus (Fig, \ref{spec_Venuscompar}).  Even though Venus is known to have a 93bar CO$_2$ atmosphere and a 730K surface temperature, its CO$_2$ absorption is significantly weaker than seen in our cases, and its  maximum brightness temperature in the MIR is only 250 K, due to its planet-wide cloud deck.

\subsection{Implications for extrasolar planet characterization missions}  

TPF-C, TPF-I, and Darwin will  all have the opportunity to study a wide variety of planetary systems.  If 
terrestrial planets are found around these stars, it is highly likely that 
we will see the atmospheres of planets, like our own Solar System examples of Venus, Mars and the early Earth, that are dominated by CO$_2$.
Therefore, understanding the appearance of CO$_2$-dominated atmospheres, 
and their ability to generate false positives is an important step in characterizing extrasolar terrestrial
planets that are not strictly Earth-like.  In addition to disproving the
likelihood of significant abiotic sources of O$_2$ 
and O$_3$ in a CO$_2$-dominated atmosphere, we have also demonstrated that CO$_2$-dominated habitable atmospheres 
exhibit a wealth of spectral features.   The multiple CO$_2$ absorption bands
throughout the visible-NIR range accessible to an optical coronograph (0.5-1.7$\mu$m) 
and an MIR interferometer (5-25$\mu$m) provide differing sensitivity to a large range of CO$_2$ abundances.   The most 
sensitive band in the visible-NIR is near 1.62$\mu$m, and produces a potentially detectable feature, even for the 
Earth's present 330ppm abundance.  For atmospheres with higher CO$_2$ abundances (0.2-2bars of CO$_2$), the less 
sensitive 1.05$\mu$m bands, which are weak in the Earth's atmosphere,  provide strong absorption features, and are 
better features for potential quantification of the planet's atmospheric
CO$_2$ provided the spectral resolution is above 50.  In the MIR, the CO$_2$ 15$\mu$m 
feature provides the strongest absorption for CO$_2$ and is the most detectable feature in a terrestrial planet 
atmospheres.   In addition, we have shown that in the MIR the isotopes of CO$_2$, produce strong features.   This is 
arguably one of the most detectable isotope features in a terrestrial planet spectrum, even at relatively poor 
spectral resolution (R $\sim 20$)  and signal to noise.   These features may allow us to get a first measure of O isotopic composition 
of an early-Earth-like or CO$_2$-dominated terrestrial planet.  As lighter isotopes tend to evaporate more readily, and 
heavier isotopes are rained out more efficiently, the atmospheric O isotope inventory may in turn provide secondary clues 
to the presence and degree of activity in a planet's hydrological cycle.

\section{Conclusions}
 Our results show that for a habitable planet with abundant liquid water, rainout of oxidized species onto a 
reduced surface will keep atmospheric H$_2$ sufficiently high, that the
generation of an abiotic ``false positive'' signal with strong O$_2$ or O$_3$
absorption is unlikely, 
and does not pose a significant hazard for habitable planets observed with TPF
or Darwin. This result is robust to realistic limits on the UV output of the
star, and the abundance of CO$_2$ in the planet's atmosphere. Dense CO$_2$
atmospheres of rocky planets may also display a range of distinctive spectral
features in both the visible and MIR that allow us to at least qualitatively 
determine the prevalence of CO$_2$ in the atmosphere, and potentially measure 
atmospheric isotopic ratios for oxygen.

\begin{acknowledgements}
This material is based upon work performed by the NASA Astrobiology 
Institute's Virtual Planetary Laboratory Lead Team, supported by the 
National Aeronautics and Space Administration through the NASA 
Astrobiology Institute under Cooperative Agreement No. CAN-00-OSS-01.  
J. F. K. also acknowledges support from NASA's Exobiology and Astrobiology Programs
M. C. thanks NASA for supporting his participation in this work
under grant NNG04GL49G with UC Berkeley. We thank Patrick Kasting for the
computational support provided. Review by F. Selsis greatly improved 
the manuscript and is acknowledged with gratitude.
\end{acknowledgements}


\end{document}